%% file: OromanaFrank.tex
\documentclass[multphys,vecphys]{svmult}

\usepackage{makeidx}     
\usepackage{graphicx}    
\usepackage{multicol}    
\usepackage{rotating}
\usepackage{cite}

\makeindex             

\begin{document}

\title*{An Introduction to Nuclear Supersymmetry: a Unification
Scheme for Nuclei}
\titlerunning{Nuclear Supersymmetry}
\author{A. Frank\inst{1}\inst{2}, J. Barea\inst{1} \and R. Bijker\inst{1}}
\institute{ICN-UNAM, AP 70-543, 04510 M\'exico, DF, M\'exico
\and CCF-UNAM, AP 139-B, 62251 Cuernavaca, Morelos, M\'exico \\
\texttt{frank@nuclecu.unam.mx},
\texttt{barea@nuclecu.unam.mx},
\texttt{bijker@nuclecu.unam.mx}}
\maketitle

\begin{abstract}
The main ideas behind nuclear supersymmetry 
are presented, starting from the basic concepts of symmetry
and the methods of group theory in physics. We propose new,
more stringent experimental tests that
probe the supersymmetry classification in nuclei and point out that
specific correlations should exist for particle transfer
intensities among supersymmetric partners. We also discuss
possible ways to generalize these ideas to cases where no
dynamical symmetries are present. The combination of these
theoretical and experimental studies may play a unifying role in
nuclear phenomena.
\end{abstract}

\section{Introduction}

One of the main objectives of research in physics is to find
simple laws that give rise to a deeper understanding and/or a 
unification of diverse phenomena. A less ambitious goal is to
construct models which, in a more or less restricted range,
permit  an understanding of the physical processes involved and
give rise  to a systematic analysis of the available experimental
data, while  providing insights into the complex systems being
studied. Among  models of nuclear structure,the Interacting Boson
Model and its extensions have proved remarkably successful in
providing a  unified framework for even-even \cite{IBM} and odd-A
nuclei \cite{IBFM}. One of its most attractive features is that it
gives rise to a simple algebraic description, where the so-called
dynamical symmetries play a central role, both as a way to
improve our basic understanding of the role of symmetry in
nuclear dynamics and as starting points from which more precise
calculations can be carried out. This approach has, in a first
stage, produced a unified description of the properties of medium
and heavy even-even nuclei, which are pictured in this framework
as belonging (in general) to transitional regions between the
dynamical symmetries. Later on, odd-A nuclei were also analyzed
using this point of view \cite{IBFM}. A further step was then taken
by  Iachello \cite{FI}, who suggested that a simultaneous
description of  even-even and odd-A nuclei was possible through
the introduction  of a superalgebra, energy levels in both nuclei
belonging to the same (super)multiplet. The idea was subsequently
tested in several  regions of the nuclear table 
\cite{susy,baha,mauthofer1,thesis,siete}. 
The step of including the odd-odd nucleus into this unifying framework was
then taken by Van Isacker et al \cite{quartet}, who managed to formulate a
supersymmetric  theory for quartets of nuclei. 

\begin{figure}[t]
\centering
\includegraphics[height=12.5cm,width=11.5cm]{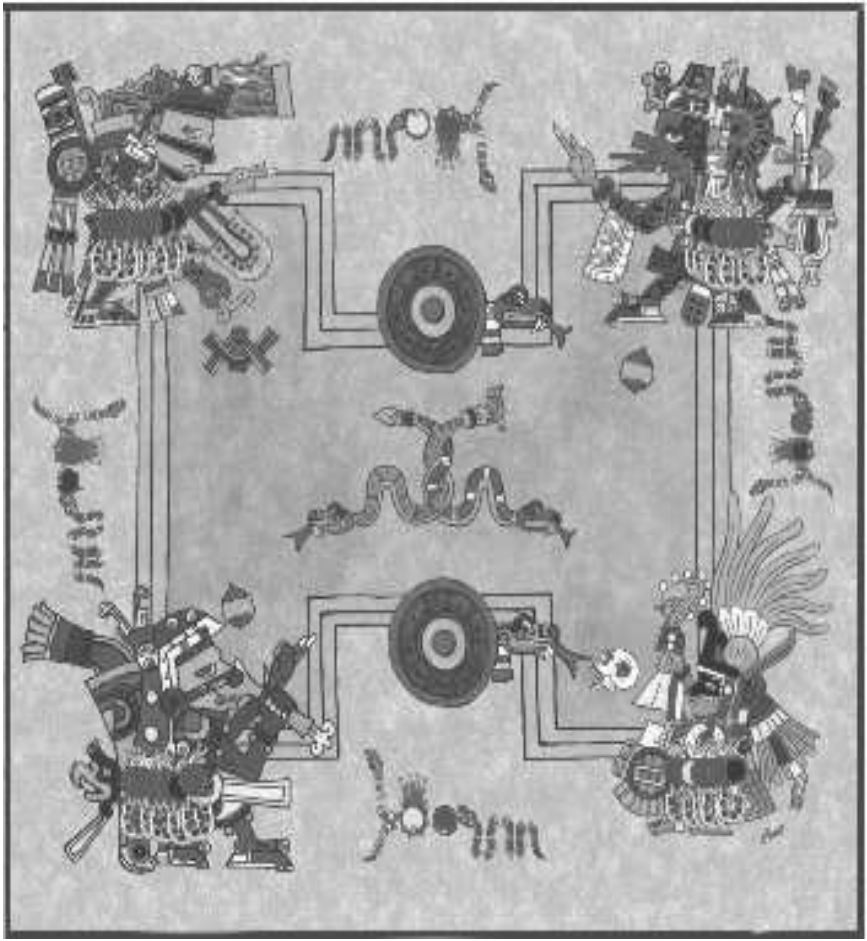}
\caption[]{\small Precolumbian Supersymmetry}
\label{aztec}
\end{figure}

In the artistic interpretation of Fig.~\ref{aztec} by Renato Lemus, 
supersymmetric quartets of nuclei are described using the language of old 
Nahua Codices (compare with Eq.~(\ref{magic}) in Section 3.4). 
Four aztec gods play the role of the supersymmetric nuclei in a quartet. 
The gods are depicted as players of the ``Juego de Pelota'', the ritual 
game of prehispanic cultures. The players carry 7 balls each which 
are color-coded. Green and blue balls correspond to neutron and proton 
bosons, while yellow and red ones to neutron and protons, respectively. 
Transfer operators are represented by coral snakes (``coralillo''), the 
traditional symbol of transformation. Creation and annihilation operators 
are identified by balls carried by the snakes, transforming one ``God'' 
into another. A more detailed explanation can be obtained from the artist 
at renato@nuclecu.unam.mx. 

In these lecture
notes we describe the basic theoretical ideas underlying nuclear
symmetry and supersymmetry (SUSY). We start these lecture notes by giving
a description of the mathematical framework needed for the
understanding of symmetries in nature, which is that of group
theory and Lie Algebras. In the subsequent sections we then
concentrate on theoretical and experimental aspects of nuclear
supersymmetry. A pedagogic description of algebraic techniques in 
nuclei and molecules can be found in \cite{nueve}, from which some of 
the following discussions have been taken. 

\section{Symmetries and Group Theory}

Symmetry and its mathematical framework---group theory---play an
increasingly important role in physics. Both classical and
quantum  systems usually display great complexity, but the
analysis of  their symmetry properties often gives rise to
simplifications and  new insights which can lead to a deeper
understanding. In  addition, symmetries themselves can point the
way toward the formulation of a correct physical theory by
providing constraints  and guidelines in an otherwise intractable
situation. It is  remarkable that, in spite of the wide variety of
systems one may  consider, all the way from classical ones to
molecules, nuclei,  and elementary particles, group theory applies
the same basic  principles and extracts the same kind of useful
information from  all of them. This universality in the
applicability of symmetry considerations is one of the most
attractive features of group  theory.  Most people have an
intuitive understanding of symmetry,  particularly in its most
obvious manifestation in terms of geometric transformations that
leave a body or system invariant. This interpretation, however, is
not enough to readily grasp its deep connections with physics, and
it thus becomes necessary to generalize the notion of symmetry
transformations to encompass more abstract ideas. The mathematical
theory of these transformations is the subject matter of group
theory. When these operations are of a continuous nature, one can
always consider the case of infinitesimal transformations and
study the behavior of the systems subject to the latter. The
mathematical theory of such transformations was first considered
by Marius Sophus Lie, who introduced the basic concepts and
operations of what are now called Lie algebras \cite{aGIL74}.

\subsection{Some Definitions}

An abstract group $G$ is
defined by a set of elements $(\hat G_1,\hat G_2,\dots,\hat G_n)$ for which
a ``multiplication'' rule combining these elements exists and
which satisfies the following conditions:
\begin{enumerate}
\item {\it Closure:} \\
If $\hat G_i$ and $\hat G_j$
are elements of the set, so is their product $\hat G_i\hat G_j$.
\item {\it Associativity:} \\
The following property is always valid:
$$\hat G_i(\hat G_j\hat G_k)=(\hat G_i\hat G_j)\hat G_k ~.$$
\item {\it Identity:} \\
There exists an element $\hat E$ of $G$ satisfying
$$\hat E\hat G_i=\hat G_i\hat E=\hat G_i ~.$$
\item {\it Inverse:} \\
For every $\hat G_i$
there exists an element  $\hat G_i^{-1}$ such that
$$\hat G_i\hat G_i^{-1}=\hat G_i^{-1}\hat G_i=\hat E ~.$$
\end{enumerate}
The number $n$ of elements is called the {\it order} of the group.
For continuous (or Lie) groups all elements may be
obtained by exponentiation in terms of a basic set of elements
$\hat g_i,i=1,2,\dots,s$, called {\it generators}, which together
form the {\it Lie algebra} associated with the Lie group. A simple
example is provided by the SO(2) group of rotations in
two-dimensional space, with elements that may be realized as
\begin{equation}
\hat G(\alpha)=e^{-i\alpha\hat l_z} ~, \label{ad1}
\end{equation}
where $\alpha$
is  the angle of rotation and
\begin{equation}
\hat l_z=-i\left(x{\frac{\partial}{\partial y}}
-y{\frac{\partial}{\partial x}} \right) ~,
\label{ad2}
\end{equation}
is the generator of these
transformations in the $x$--$y$ plane. Three-dimensional
rotations  require the introduction of two additional generators,
associated with rotations in the $z$--$x$ and $y$--$z$ planes,
\begin{equation}
 \hat l_y=-i\left(z{\frac{\partial}{\partial x}}
-x{\frac{\partial}{\partial z}}\right)~, \qquad
 \hat l_x=-i\left(y{\frac{\partial}{\partial z}}
-z{\frac{\partial}{\partial y}}\right) ~,
\label{ad3}
\end{equation}
Finite rotations can then be parametrized by
three angles (which may be chosen to be the Euler angles) and
expressed as a product of exponentials of the so(3) generators
(\ref{ad2}) and (\ref{ad3}) \cite{aROS57}. Evaluating the
commutators of these operators, we find
\begin{equation} [\hat l_x,\hat l_y]=i\hat l_z~,
\qquad [\hat l_y,\hat l_z]=i\hat l_x~, 
\qquad [\hat l_z,\hat l_x]=i\hat l_y~,
\label{ad4}
\end{equation}
which illustrates the closure property of the
group generators. In general, the $s$ operators
$\hat g_i,i=1,2,\dots,s$, define a {\it Lie algebra} if they close
under commutation,
\begin{equation} [\hat g_i,\hat g_j]=\sum_k c^k_{ij}\hat g_k ~,
\label{ad5}
\end{equation}
and satisfy the Jacobi identity \cite{aWYB74}
\begin{equation}
[\hat g_i,[\hat g_j,\hat g_k]]+[\hat g_k,[\hat g_i,\hat g_j]]
+[\hat g_j,[\hat g_k,\hat g_i]]=0~.
\label{ad6}
\end{equation}
The set of constants $c^k_{ij}$ are called
{\it structure constants}, and their values determine the
properties of both the Lie algebra and its associated Lie group.
All Lie groups have been classified by Cartan
\cite{aWYB74,aLIP66}, and many of their properties have
been established.

\subsection{Symmetry Transformations}

From a general point of view symmetry transformations of a
physical system may be defined in terms of the equations of motion
for the system \cite{aMAN91}. Suppose we consider the system of
equations
\begin{equation} {\cal O}_i \psi_i({\bf x})=0~, \qquad i=1,2,\dots~,
\label{at1}
\end{equation}
where the functions $\psi_i({\bf x})$ denote a vector column with a
finite or infinite number of components, or a more general
structure such as a matrix depending on the variables $x_i$. The
operators $\cal{O}_i$ are quite arbitrary, and (\ref{at1}) may
correspond, for example, to Maxwell, Schr\"odinger, or Dirac
equations. The operators $\hat g_{ij}$ such that
\begin{equation}
\sum_j {\cal O}_i(\hat g_{ij} \psi_j)=0, \qquad i=1,2,\dots~,
\end{equation}
are called symmetry transformations, since they transform
 the solutions $\psi$ to other solutions $g \psi$ of the equations
 (\ref{at1}). As a particular example we consider the
 time-dependent Schr\"odinger equation (with $\hbar=1$)
\begin{equation}
 \left(\hat H({\bf x},{\bf p})-i\frac{\partial}{\partial t}\right)
\psi({\bf x},t)=0~.
 \label{at3}
\end{equation}
One can verify that
$\hat k_j({\bf x},{\bf p},t)\psi({\bf x},t)$ is also a solution of
(\ref{at3}) as long as $\hat k_j$ satisfies the equation
\begin{equation}
[\hat H,\hat k_j]-i{{\partial\hat k_j}\over{\partial t}}=0~,
\label{at4}
\end{equation}
 which means that $\hat k_j$ is an operator associated with a
conserved quantity. The last statement follows from the definition
 of the total derivative of an operator $\hat A_j$
\begin{equation}
{{d\hat A_j}\over{dt}} ={{\partial\hat A_j}\over{\partial t}}
+i[\hat H,\hat A_j]~,
\end{equation}
where $\hat H$ is the quantum-mechanical Hamiltonian
\cite{aMES68}. If $\hat k_1$ and $\hat k_2$ satisfy (\ref{at4}), their
 commutator is again a constant of the motion since
\begin{eqnarray}
\frac{d}{dt} [\hat k_1,\hat k_2] &=&
\frac{\partial}{\partial t}[\hat k_1,\hat k_2]+i[\hat H,[\hat k_1,\hat k_2]]
 \nonumber\\
&=& \frac{\partial}{\partial t} [\hat k_1,\hat k_2]
-[{{\partial\hat k_1}\over{\partial t}},\hat k_2]
-[\hat k_1,{{\partial\hat k_2}\over{\partial t}}]=0~,
\end{eqnarray}
where use is made of (\ref{at4}) and the Jacobi identity
 (\ref{ad6}). A particularly interesting situation arises when the
 set $(\hat k_i)$ is such that $[\hat k_i,\hat k_j]$ closes under
 commutation to form a Lie algebra as in (\ref{ad5}). In this case
 we refer to $(\hat k_i)$ as the generators of the {\em symmetry}
 (Lie) algebra of the time-dependent quantum system (\ref{at3})
 \cite{aCAS90}. Note that in general these operators do not
 commute with the Hamiltonian but rather satisfy (\ref{at4}),
\begin{equation}
 [\hat H-i\frac{\partial}{\partial t},\hat k_j]=0~.
\end{equation}
What about the time-independent Schr\"odinger equation? This
 case corresponds to substituting
 $\psi({\bf x},t)=\psi_n(x)e^{-iE_nt}$ in (\ref{at3}), leading to
 \begin{equation}
(\hat H({\bf x},{\bf p})-E_n)\psi_n({\bf x})=0~.
\label{at8}
\end{equation}
The set $\hat k_j({\bf x},{\bf p},t=0)$ still satisfies the same commutation
 relations as before but due to (\ref{at4}) are not in general
integrals of the motion anymore. These operators constitute the
{\it dynamical algebra} for the time-independent Schr\"odinger
 equation (\ref{at8}) and connect all solutions $\psi_n(x)$ with
 each other, including states at different energies. Due again to
 (\ref{at4}), only those $\hat k_j$ generators that are time
 independent satisfy
\begin{equation}
[\hat H,\hat k_j]=0~,
\label{at9}
\end{equation}
which
 implies that they are constants of the motion for the system
 (\ref{at8}). Equation (\ref{at9}) (together with the closure of
 the $\hat k_j$'s) constitutes the familiar definition of the symmetry
 algebra for a time-independent system. The connection between the
 dynamical algebra $(\hat k_j(0))$ and the symmetry algebra of the
 corresponding time-dependent system $(\hat k_j(t))$ allows a unique
 definition of the dynamical algebra \cite{aCAS90}.

 \subsection{Constants of the Motion and State Labeling}
 \label{constants}

 From the previous discussion we see
 that the symmetry Lie algebras associated with both the
 time-dependent and time-independent Schr\"o\-dinger equations
 supply integrals of the motion for physical systems. In addition,
 the dynamical algebra of the latter is such that all solutions
 $\psi_n({\bf x})$ are connected by means of its generators. This
 means that the dynamical algebra implicitly defines the
 appropriate Hilbert space for the description of the physical
 system. For any Lie algebra one may construct one or more
 operators ${\cal C}_l$ which commute with all the generators $\hat k_j$,
\begin{equation}
[{\cal C}_l,\hat k_j]=0, \qquad l=1,2,\dots,r, \quad j=1,2,\dots,s~.
 \end{equation}
These operators are called {\it Casimir operators} or {\it
 Casimir invariants}, and there are many examples for the u($n$)
 and so($n$) algebras of the kind we discuss later on. 
They may be linear, quadratic, or of higher
 order in the generators. The number $r$ of linearly independent
 Casimir operators is called the rank of the algebra
 \cite{aWYB74}. This number coincides with the maximum subset of
 generators which commute among themselves (called {\it weight
 generators})
\begin{equation}
[\hat k_\alpha,\hat k_\beta]=0~, \qquad \alpha,\beta=1,2,\dots,r~,
\label{ac2}
\end{equation}
where we use greek
 labels to indicate that they belong to the subset satisfying
 (\ref{ac2}). The operators $({\cal C}_i,\hat k_\alpha)$ may be
 simultaneously diagonalized and their eigenvalues used to label
 the corresponding eigenstates.

To illustrate these definitions, we consider the su(2) algebra
$(\hat j_x,\hat j_y,\hat j_z)$ with commutation relations
\begin{equation}
 [\hat j_x,\hat j_y]=i\hat j_z~, \qquad [\hat j_z,\hat j_x]=i\hat j_y~, \qquad
 [\hat j_y,\hat j_z]=i\hat j_x~, \label{ac3}
\end{equation}
isomorphic to the so(3) commutators given in (\ref{ad4}).
 From (\ref{ac3}) we conclude  that $r=1$
 and we may choose $\hat j_z$ as the generator to diagonalize together
 with the Casimir invariant
\begin{equation}
\hat j^2=\hat j^2_x+\hat j^2_y+\hat j^2_z~.
 \end{equation}
The eigenvalues and branching rules for the commuting set
 $({\cal C}_l,\hat k_\alpha)$ can be determined solely from the commutation
 relations (\ref{ad5}). In the case of su(2) the eigenvalue
 equations are
\begin{equation}
\hat j^2\vert jm\rangle=n_j\vert jm\rangle~,
 \qquad \hat j_z\vert jm\rangle=m\vert jm\rangle~,
 \end{equation}
where $j$ is an index to distinguish the different $\hat j^2$
 eigenvalues. Defining the raising and lowering operators
\begin{equation}
 \hat j_\pm=\hat j_x\pm i\hat j_y~,
 \end{equation}
and using (\ref{ac3}), one finds the well-known results \cite{aROS57}
\begin{equation}
n_j=j(j+1)~, \quad j=0,1/2,1,\dots,
 \quad m=-j,-j+1,\dots,j~.
 \end{equation}
As a bonus, the action of $\hat j_\pm$ on the $\vert jm\rangle$
 eigenstates is also determined to be
\begin{equation}
\hat j_\pm \vert jm\rangle = \sqrt{(j\mp m)(j\pm m+1)} \vert jm\pm 1 \rangle~.
 \end{equation}
In the case of a general Lie algebra (\ref{ad5}) the
 procedure can be quite complicated but requires the same basic
 steps. The analysis leads to the algebraic determination of
 eigenvalues, branching rules, and matrix elements of raising and
 lowering operators \cite{aWYB74}.

 Returning to the time-independent Schr\"odinger equation, it
 follows from our discussion that the {\em symmetry} algebra
 provides constants of the motion, which in turn lead to quantum
 numbers that label the states associated with a given energy
 eigenvalue. The raising and lowering operators in this algebra
 only connect degenerate states. The dynamical algebra, however,
 defines the whole set of eigenstates associated with a given
 system. The generators are no longer constants of the motion as
 not all commute with the Hamiltonian. The raising and lowering
 operators may now connect all states with each other.

 \subsection{Eigenfunctions and representations}

 For a given group $G$ of physical operations $(\hat R)$ one may
 introduce a set of operators $\hat P_R$ which are defined by their
 action on an arbitrary scalar function $f({\bf x})$:
\begin{equation}
 \hat P_Rf({\bf x})=f(\hat R{\bf x})~.
\label{ae1}
\end{equation}
The correspondence
$\hat R\rightarrow\hat P_R$ is an isomorphism, as
$\hat S\hat R\rightarrow\hat P_S\hat P_R=\hat P_{SR}$, as can be shown from
(\ref{ae1}). A simple example is provided by the two-dimensional
 rotations (\ref{ad1}). To deduce their explicit form we apply
 (\ref{ae1}), using polar coordinates
\begin{equation}
\hat P_\alpha
f(r,\phi)=f(r,\phi-\alpha)~,
 \end{equation}
expand in a Taylor series,
\begin{eqnarray}
f(r,\phi-\alpha) &=& \sum_{n = 0}^\infty (-\alpha)^n {1\over{n!}}
 {{\partial^n f(r,\phi)}\over{\partial\phi^n}}
\nonumber\\
&=& \sum_{n = 0}^\infty {1\over{n!}}
\left(-\alpha\frac{\partial}{\partial \phi}\right)^nf(r,\phi)
\nonumber\\
&=& e^{-\alpha\partial/\partial\phi}f(r,\phi)~,
 \end{eqnarray}
leading to
\begin{equation}
\hat P_\alpha=e^{-i\alpha\hat l_z}~, \qquad \hat l_z
 =-i\frac{\partial}{\partial \phi}
 =-i\left(x\frac{\partial}{\partial y}-y\frac{\partial}{\partial x}\right)~,
 \end{equation}
which coincides with (\ref{ad1}) and (\ref{ad2}).

 Now consider the defining equation
\begin{equation}
 \hat H({\bf x})f({\bf x})=g({\bf x})~,
 \end{equation}
where $\hat H({\bf x})$ is an operator. Using this definition and
 the property (\ref{ae1}), we find the following two relations:
 \begin{equation}
 \begin{array}{l}
 \hat P_R\hat H({\bf x})\hat P_R^{-1}\hat P_Rf({\bf x})
=\hat P_Rg({\bf x}) =g(\hat R{\bf x})
 =\hat H(\hat R{\bf x})f(\hat R{\bf x})~,
 \nonumber\\
 \hat P_R\hat H({\bf x})\hat P_R^{-1}\hat P_Rf({\bf x})
 =\hat P_R\hat H({\bf x})\hat P_R^{-1}f(\hat R{\bf x})~.
 \end{array}
 \end{equation}
Since $f({\bf x})$ is an arbitrary function, comparison of the
 right-hand sides of these equations shows that operators
transform  as
\begin{equation}
\hat P_R\hat H({\bf x}) \hat P_R^{-1}=\hat H(\hat R{\bf x})~.
 \end{equation}
If for all $\hat R$ we have
\begin{equation}
 \hat P_R\hat H({\bf x})\hat P_R^{-1}=\hat H({\bf x})~,
\label{ae8}
\end{equation}
then
 $\hat H({\bf x})$ is said to be invariant under the action of the group
 $G=(\hat R)$ or that $G$ is a symmetry group for $H({\bf x})$. This
 definition coincides with our general discussion leading to
 (\ref{at9}), as (\ref{ae8}) implies
\begin{equation}
[\hat P_R,\hat H({\bf x})]=0~.
 \label{ae9}
\end{equation}
Let us return to the time-independent
 Schr\"odinger equation
\begin{equation}
\hat H\psi=E\psi~,
 \end{equation}
and use (\ref{ae9}).
We find \begin{equation}
\hat H(\hat P_R\psi)=E(\hat P_R\psi)~.
\label{ae11}
\end{equation}
Suppose that the
 eigenvalue $E$ is degenerate and that $l$ independent
eigenfunctions $\psi_1,\psi_2,\dots,\psi_l$ are associated with
 it. Since (\ref{ae11}) implies that $\hat P_R\psi$ is also an
 eigenfunction of $\hat H$ associated with $E$, it must be a
linear combination of the $\psi_i$s,
\begin{equation}
\hat P_R\psi_i({\bf x})
=\sum_{j=1}^l D_{ji}(\hat R) \psi_j ({\bf x})~, \qquad i=1,2,\dots,l~.
 \end{equation}
The matrices $D_{ji}(\hat R)$ are called a {\it representation}
 of the group $G$, and it is easy to prove that they satisfy the
 matrix product
\begin{equation} D(\hat S)D(\hat R)=D(\hat S\hat R)~.
 \end{equation}
The $l$ independent eigenfunctions
 $\psi_1,\psi_2,\dots,\psi_l$ are said to constitute a basis for
 this representation. In addition, if the $\psi_i$'s are such that
 no change of basis transformation
\begin{equation}
\phi_i=\sum_jU_{ij}\psi_j~,
 \end{equation}
can take all the ${\bf D}$ matrices to block-diagonal form,
 that is, to the form
\begin{equation}
\bf U^{-1}\bf D \bf U \rightarrow
 \left[\begin{array}{ccc}
 {\bf D_1}&\vdots&{\bf 0}\\
 \cdots&\cdot&\cdots\\
 {\bf 0}&\vdots&{\bf D_2}
 \end{array}\right]~,
 \label{ae15}
\end{equation}
we then say that the representation is {\it
 irreducible} and that the $\psi_i$'s are a basis for an {\it
 irreducible representation} of $G$. The form (\ref{ae15}) would
 imply that two subsets of the $l$ $\psi_i$'s transform only among
 themselves under the action of $G=(\hat R)$.

 As an example we return to the SO(3) group where the appropriate
 basis for the irreducible representations is given by the
 spherical harmonics \cite{aROS57} $Y^l_m(\theta,\phi)$. The
 action of the rotation-group elements gives
\begin{equation}
 \hat P_R(\theta_1,\theta_2,\theta_3) Y^l_m(\theta,\phi) =\sum_{m'}
 D^l_{m'm}(\theta_1,\theta_2,\theta_3) Y^l_{m'}(\theta,\phi)~,
 \label{ae16}
\end{equation}
where Wigner's ${\bf D}$ matrices are introduced
 \cite{aROS57}, which play the role of SO(3) irreducible
 representations. We further note that the $Y_{lm}(\theta,\phi)$
 satisfy the eigenvalue equations
\begin{equation}
 \hat l^2Y^l_m(\theta,\phi)=l(l+1)Y^l_m(\theta,\phi)~, \qquad
 \hat l_zY^l_m(\theta,\phi)=mY^l_m(\theta,\phi) ~, 
\label{ae17}
\end{equation}
 where $\hat l^2$ is the SO(3) Casimir invariant
\begin{equation}
 \hat l^2=\hat l^2_x+\hat l^2_y+\hat l^2_z~.
 \end{equation}
This symmetry group (and its algebra) applies for all
 Hamiltonians invariant under physical rotations. For
arbitrary Lie groups relation (\ref{ae16}) is generalized to
\begin{equation}
 \hat P_Rf^\lambda_\mu({\bf x}) =\sum_{\mu'} \
D^\lambda_{\mu'\mu}(\hat R) f^\lambda_{\mu'}({\bf x})~,
\label{ae19}
\end{equation}
where $\lambda$ denotes in general a set of quantum numbers that
label the irreducible representations of the group $G=(\hat R)$ and
$\mu$ (and $\mu'$) label the different functions in the
representation. They are often chosen to correspond to sets of
quantum numbers that label the irreducible representations of
subgroups of $G$. Likewise, (\ref{ae17}) is generalized to
\begin{equation}
{\cal C}_lf^\lambda_\mu({\bf x}) =h_l(\lambda)f^\lambda_\mu({\bf x})~,
\qquad \hat k_\alpha f^\lambda_\mu({\bf x})
=h_\alpha(\mu)f^\lambda_\mu({\bf x})~,
\label{ae20}
\end{equation}
where ${\cal C}_l$
and $\hat k_\alpha$ are the Casimir invariants and weight generators
defined in Subsection~\ref{constants}. The eigenvalues $h_l(\lambda)$ and
$h_\alpha(\mu)$ may be determined from the commutation relations
that define the Lie algebra associated with $G$, as explained in
the previous section.

\subsection{The Algebraic Approach}
\label{algebraic}

 In this section we
show how the concepts presented in the previous sections lead to
an algebraic approach which can be applied to the study of different 
physical systems.
 We start by considering again (\ref{at9}), which
describes the invariance of a Hamiltonian under the algebra
$g=(\hat k_j)$,
\begin{equation} [\hat H,\hat k_j]=0~,
\label{aa1}
\end{equation}
implying that $g$
plays the role of symmetry algebra for the system. Equation
(\ref{ae11}), on the other hand, implies that an eigenstate of
$\hat H$ with energy $E$ may be written as $\vert\lambda\mu\rangle$,
where $\lambda$ labels the irreducible representations of the
group $G$ corresponding to $g$ and $\mu$ distinguishes between the
different eigenstates with energy $E$ (and may be chosen to
correspond to irreducible representations of subgroups of $G$).
The energy eigenvalues of the Hamiltonian in (\ref{aa1}) thus
depend only on $\lambda$,
\begin{equation}
\hat H\vert\lambda\mu\rangle
=E(\lambda)\vert\lambda\mu\rangle~,
 \end{equation}
and furthermore, (\ref{ae19}) implies that the generators
 $\hat k_i$ (and their corresponding group operators $\hat P_R$) do not
 admix states with different $\lambda$'s.
 The use of the mutually commuting set of Casimir invariants and
 generators described in the previous section then leads to the
 full specification of the states $\vert\lambda\mu\rangle$
 through (\ref{ae20}).

 We now consider the chain of algebras
\begin{equation} g_1\supset g_2~,
 \end{equation}
which will lead us to introduce the concept of {\it
 dynamical symmetry}. If $g_1$ is a symmetry algebra for $\hat H$, we
 may label its eigenstates as $\vert\lambda_1\mu_1\rangle$. Since
 $g_2 \subset g_1$, $g_2$ must also be a symmetry algebra for $\hat H$
 and, consequently, its eigenvalues labeled as
 $\vert\lambda_2\mu_2\rangle$.  Combination of the two properties
 leads to the eigenequation
\begin{equation}
 \hat H\vert\lambda_1\lambda_2\mu_2\rangle
 =E(\lambda_1)\vert\lambda_1\lambda_2\mu_2\rangle~,
\label{aa4}
\end{equation}
 where the role of $\mu_1$ is played by $\lambda_2\mu_2$ and hence
 the eigenvalues depend only on $\lambda_1$. This process may be
 continued when there are further subalgebras, that is, $g_1\supset
 g_2\supset g_3\supset\cdots$, in which case $\mu_2$ is substituted
 by $\lambda_3\mu_3$, and so on.

 In many physical applications the original assumption that $g_1$
 is a symmetry algebra of the Hamiltonian is found to be too strong
 and must be relaxed, that is, one is led to consider the breaking
 of this symmetry. An elegant way to do so is by considering a
 Hamiltonian of the form
\begin{equation} \hat H'=a {\cal C}_{l_1}(g_1)+b {\cal C}_{l_2}(g_2)~,
 \label{aa5}
\end{equation}
where ${\cal C}_{l_i}(g_i)$ is a Casimir invariant of
 $g_i$. Since $[\hat H',\hat k_i]=0$ for $\hat k_i\in g_2$, $\hat H'$ is
 invariant under $g_2$, but not anymore under $g_1$ because
 $[{\cal C}_{l_2}(g_2),\hat k_i]\neq 0$ for $\hat k_i\not\in g_2$.
The new {\it symmetry algebra} is thus $g_2$ while $g_1$ now plays the role of
 {\it dynamical algebra} for the system, as long as all states we
 wish to describe are those originally associated with
 $E(\lambda_1)$. The extent of the symmetry breaking depends on the
 ratio $b/a$. Furthermore, since $\hat H'$ is given as a combination
 of Casimir operators, its eigenvalues can be obtained in closed
 form using (\ref{ae20}):
\begin{equation}
 \hat H'\vert\lambda_1\lambda_2\mu_2\rangle
 =(aE_{l_1}(\lambda_1)+bE_{l_2}(\lambda_2))
 \vert\lambda_1\lambda_2\mu_2\rangle~.
\label{aa6}
\end{equation}
The kind of
 symmetry breaking caused by interactions of the form (\ref{aa5})
 is known as {\it dynamical-symmetry breaking} and the remaining
 symmetry is called a {\it dynamical symmetry} of the Hamiltonian
 $\hat H'$.
 From (\ref{aa6}) we conclude that
 even if $\hat H'$ is not invariant under $g_1$, its eigenstates are
 the same as those of $\hat H$ in (\ref{aa4}). The dynamical-symmetry
 breaking thus splits but does not admix the eigenstates.

 The algebraic approach often makes use of dynamical symmetries to
 compute energy eigenvalues, but it goes further in order to
 describe all relevant aspects of a system in purely algebraic
 terms. To do so, it follows a number of steps:
 \begin{enumerate}
 \item A given system is described in terms of a dynamical algebra
 $g_1$ which spans all possible states in the system within a fixed
 irreducible representation. The choice of this algebra is often
 dictated by physical considerations (such as the quadrupole nature
 of collective nuclear excitations or the dipole character of
 diatomic molecular vibrations).
\item The Hamiltonian and all
 other operators in the system, such as electromagnetic multipole
 operators, should be expressed entirely in terms of the generators
 of the dynamical algebra. Since the matrix elements of the
 generators can be evaluated from the commutation properties of the
 dynamical algebra, this implies that all observables of the system
 can be calculated algebraically.
\item The appropriate bases for
 the computation of matrix elements are supplied by the different
 dynamical symmetries associated with the Hamiltonian. Physically
 meaningful chains are those where the symmetry algebra of the
 Hamiltonian is a subalgebra of the dynamical algebra in the chain
 $g_1\supset g_2\supset\cdots$ chosen to label these bases.
\item
 Branching rules for the different algebra chains as well as
 eigenvalues of their Casimir operators need to be evaluated to
 fully determine the dynamical symmetry bases and their associated
 energy eigenvalues.
\item When several dynamical symmetry chains
 containing the symmetry algebra are present in the system, the
 Hamiltonian will in general not be diagonal in any given chain but
 rather include invariant operators of all possible subalgebras. In
 that case the Hamiltonian should be diagonalized in one of these
 bases. Dynamical symmetries are still useful as limiting cases
 where all observables can be analytically determined.
 \end{enumerate}

 We remark that the condition 1, namely that all states of the
 system should be spanned by a single irreducible representation of
 the dynamical algebra $g_1$, assures that all states of the system
 can be reached by means of the generators of $g_1$. If this
 condition is not satisfied (e.g., if two or more irreducible
 representations would span the states), step 2 indicates that the
 physical operators would not connect the states in different
 irreducible representations and would constitute independent sets.

\begin{figure}[t]
centering \includegraphics[height=7cm]{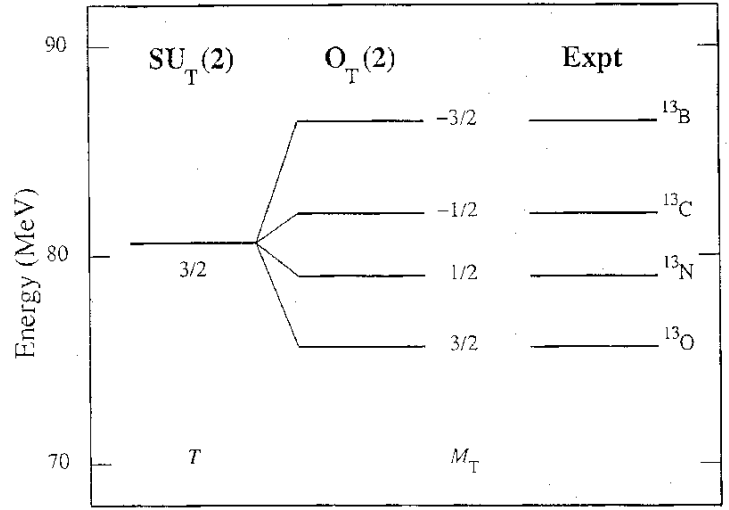} 
\caption[]{\small
Binding energies of the $T=3/2$ isobaric analog states with
angular momentum and parity $J^\pi=1/2^-$ in $^{13}$B,
$^{13}$C, $^{13}$N, and $^{13}$O. The column on the left is
obtained for an exact $SU_T(2)$ {\em symmetry}, which predicts
states with different $M_T$ to be degenerate. The middle column is
obtained in the case of an $SU_T(2)$ {\em dynamical symmetry},
 equation (\ref{aa7}) with parameters $a=80.59$, $b=-2.96$, and
 $c=-0.26$ MeV.}
\label{isobaric}
\end{figure}

 Some of these ideas can be illustrated with well-known examples.
 In 1932 Heisenberg considered the occurrence of isospin multiplets
 in nuclei \cite{aHEI32}. To a first approximation neutrons and
 protons in nuclei interact through isospin-invariant forces, that
 is, to this approximation the electromagnetic effects are
 neglected compared with the strong interaction. In the notation
used above (without making the distinction between algebras and
 groups), $G_1$ is in this case the isospin algebra $SU_T(2)$,
 consisting of the operators $\hat T_x$, $\hat T_y$,
and $\hat T_z$ which
 satisfy commutation relations (\ref{ac3}),
and $G_2$ can be identified with $SO_T(2)=(\hat T_z)$. An
isospin-invariant Hamiltonian commutes with $\hat T_x$, $\hat T_y$, and
$\hat T_z$, and hence the eigenstates $\vert TM_T\rangle$ with fixed
$T$ and $M_T=-T,-T+1,\dots,T$ are degenerate in energy. The next
approximation is to take into account the electromagnetic
interaction which breaks isospin invariance and lifts the
degeneracy of the states $\vert TM_T\rangle$. It is assumed that
this symmetry breaking occurs dynamically, and since the Coulomb
force has a two-body character, the breaking terms are at most
quadratic in $\hat T_z$ \cite{aELL79}. The energies of the
corresponding nuclear states with the same $T$ are then given by
\begin{equation} 
E(M_T)=a+bM_T+cM_T^2~,
\label{aa7}
\end{equation}
and $SU_T(2)$
becomes the dynamical symmetry for the system while $SO_T(2)$ is
the symmetry algebra. The dynamical symmetry breaking thus implied
that the eigenstates of the nuclear Hamiltonian have well-defined
values of $T$ and $M_T$. Extensive tests have shown that indeed
this is the case to a good approximation, at least at low
excitation energies and in light nuclei \cite{aBOH75}. Formula
(\ref{aa7}) can be tested in a number of cases. In
Figure~\ref{isobaric} a
$T=3/2$ multiplet consisting of states in $^{13}$B,
$^{13}$C, $^{13}$N, and $^{13}$O is compared with the theoretical
 prediction (\ref{aa7}).

 A less trivial example of dynamical-symmetry breaking is provided
 by the Gell-Mann--Okubo mass-splitting formula for elementary
 particles \cite{aGEL62,aOKU62}. The SU(3) model of
 Gell-Mann and Ne'eman \cite{aGEL64} classifies hadrons as SU(3)
 multiplets, that is, a given irreducible representation
 $(\lambda,\mu)$ of SU(3) of dimension $d$ contains $d$ particles.
 For example, the neutron and proton are placed in the eight
 dimensional representation (1,1), the so-called {\it octet}
 representation. Besides isospin $T$ a new quantum number is needed
 to fully classify the SU(3) states. This turns out to be an
 additive number $Y$, called {\it hypercharge} \cite{aELL79},
 associated with the chain of algebras
\begin{equation}
 \begin{array}{ccccccccc}
 SU(3)&\supset&U_Y(1)&\otimes&SU_T(2)&\supset&U_Y(1)&\otimes&SO_T(2)\\
 \vert&&\vert&&\vert&&&&\vert\\
 (\lambda,\mu)&&Y&&T&&&&M_T
 \end{array}
 \label{aa8}
\end{equation}
If one would assume SU(3) invariance, all
 particles in a multiplet would have the same mass, but since the
 experimental masses of other baryons differ from the nucleon
 masses by hundreds of MeV, the SU(3) symmetry clearly must be
 broken.

\begin{figure}[t]
\centering 
\includegraphics[height=7cm]{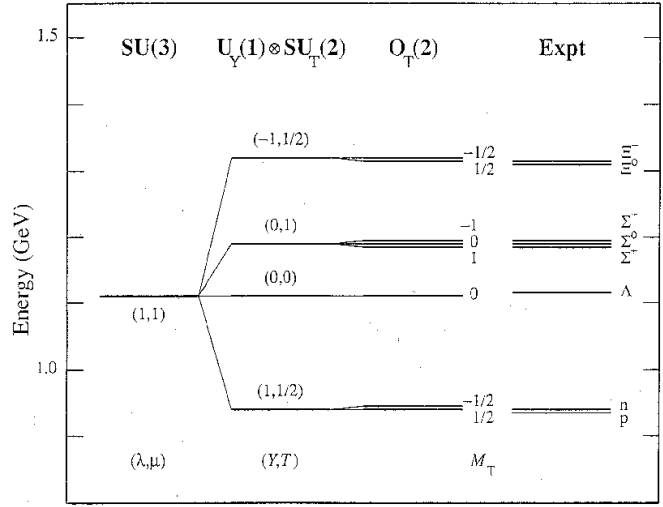} 
\caption{\small Mass spectrum of the SU(3)
octet $(\lambda,\mu)=(1,1)$. The column on the left is obtained
 for an exact SU(3) {\em symmetry}, which predicts all masses
to be the same, while the next two columns represent successive
 breakings of this symmetry in a dynamical manner. The column
under $SO_T(2)$ is obtained with (\ref{okubo}) with parameters 
$a=1111.3$, $b=-189.6$, $d=-39.9$, $e=-3.8$, and
$f=0.9$ MeV.}\label{octete}
\end{figure}

 Dynamical symmetry breaking allows the baryon states to still be
 classified by (\ref{aa8}). Following the procedure outlined above
 and keeping up to quadratic terms, one finds a mass operator of
 the form
\begin{eqnarray}
 \hat M&=&a+b \, {\cal C}_{1U_Y(1)}+c \, {\cal C}^2_{1U_Y(1)}
+d \, {\cal C}_{2SU_T(2)} +e \, {\cal C}_{1SO_T(2)} 
+f \, {\cal C}^2_{1SO_T(2)} ~,
\end{eqnarray}
with eigenvalues
\begin{equation}
M(Y,T)=a+b\,Y+c\,Y^2+d\,T(T+1)+e\,M_T+f\,M_T^2~.
\label{aa10}
\end{equation}
A further assumption regarding the SU(3) tensor character of the
 strong interaction \cite{aELL79,aGEL62,aOKU62}
leads to a relation between $c$ and $d$ in (\ref{aa10}), resulting
in the Gell-Mann-Okubo mass formula
\begin{equation}
M'(Y,T)=a+bY+d\left[T(T+1)-\frac{1}{4}Y^2 \right]~.
\label{okubo}
 \end{equation}
In Figure~\ref{octete} this process of successive dynamical-symmetry
 breaking is illustrated with the octet representation containing
 the neutron and the proton and the $\Lambda$, $\Sigma$, and $\Xi$
 baryons.
Other hadrons are analogously classified using SU(3) as the
dynamical algebra \cite{aELL79,aGEL64}. Other applications of the
 algebraic approach will be illustrated throughout these lecture notes,
where the steps listed before are implemented for physical systems
associated with $U(n)$ models. The algebraic
approach, both in the sense we have defined here and in its
generalizations to other fields of research, has become an
important tool in the search for a unified description of physical
phenomena. This is illustrated by Figure~\ref{octete}. The near equality of
the neutron and proton masses suggested the existence of isospin
multiplets, later confirmed at higher energies for other
particles. To find a relationship between these multiplets, the
$SU(3)$ dynamical algebra was proposed (and became the basis
for the establishment of the quark model). This unification
process can be continued: different $(\lambda,\mu)$ multiplets can
be unified by means of higher-dimensional algebras such as $SU(4)$
\cite{aELL79}.

\subsection{Superalgebras}

\input{mac.tex}
To conclude the mathematical introduction, we now introduce the
concept of superalgebra, which generalizes the  algebras 
discussed in the previous sections and which is intimately related to the
supersymmetry concept.

The mathematical structures based on the better known (Lie) algebras 
introduced before
  can be realized in terms of either bosons or fermions. A
simple way to do this is to consider a system of bosons (fermions)
which can be in $n$ ($m$) different states denoted by $\alpha,
\alpha',...$ ($\beta, \beta',...$). They can be created or annhilated
by the
creation and annhilation operators $\db_\alpha$ ($\da_\beta$) and
$b_\alpha$ ($a_\beta$), which satisfy
\beqn
\begin{array}{lcl}
[\,b_\alpha, \db_{\alpha'} ] & = & \delta_{\alpha \alpha'} ~,
\\[1ex]
\{a_\beta, \da_{\beta'} \} & = & \delta_{\beta \beta'}~,
\end{array}
\eeqn
with all other commutators (anticommutators) being zero. The bilinear
products $(\cG^\rB_\rB)_{\alpha \alpha'}=\db_\alpha b_{\alpha'}$ ($(
\cG^\rF_\rF )_{\beta \beta'}=\da_\beta a_{\beta'}$) can be shown to 
close under commutation
\beqn
\begin{array}{lcl}
[(\cG^\rB_\rB)_{\alpha \alpha'},(\cG^\rB_\rB)_{\alpha'' \alpha'''} ] 
 & = &
(\cG^\rB_\rB)_{\alpha \alpha''} \delta_{\alpha' \alpha'''} - 
(\cG^\rB_\rB)_{\alpha' \alpha'''} \delta_{\alpha \alpha''} ~,
\\[1ex]
[(\cG^\rF_\rF)_{\beta \beta'},(\cG^\rF_\rF)_{\beta'' \beta'''} ] 
 & = &
(\cG^\rF_\rF)_{\beta \beta''} \delta_{\beta' \beta'''} - 
(\cG^\rF_\rF)_{\beta' \beta'''} \delta_{\beta \beta''}~,
\end{array}
\eeqn
and satisfy the Jacobi identity. So they define a Lie algebra of the
general $\rU^\rB(n)$ ($\rU^\rF(m)$) form. Since boson and fermion 
operators commute
\beqn
[(\cG^\rB_\rB)_{\alpha \alpha'},(\cG^\rF_\rF)_{\beta \beta'}]=0~.
\eeqn
the set of operators ($(\cG^\rB_\rB)_{\alpha \alpha'},
(\cG^\rF_\rF)_{\beta \beta'}$) define the direct-product algebra
\beqn
\rU^\rB(n)\otimes\rU^\rF(m)~,
\eeqn
which is the dynamical algebra for the combined boson-fermion system 
(see Section 3.2).

The Hamiltonian of the boson, fermion or boson-fermion system 
can be built in terms of the bilinear products or generators
of the corresponding dynamical algebras and separately conserves
the boson and fermion numbers.  The question arises as to  whether
one may define a generalized dynamical algebra  where cross terms of
the type $\db_\alpha a_{\beta}$ or $\da_{\beta} b_\alpha $ 
are included  and, if so, to study the consequences of this
generalization. From the standpoint of fundamental processes,  where
bosons correspond to forces  (i.e. photons, gluons, etc.)  and
fermions to matter (i.e. electrons, nucleons, quarks, etc.),  it may
seem strange at first sight to consider symmetries  which mix such
intrinsically different particles.  However, there have been numerous 
applications of these ideas in the last few years.
These symmetries---known as supersymmetries---have  given rise to schemes
which hold promise in quantum field theory  in regards to the
unification of the fundamental interactions 
\cite{4WES74,4FAY77,4NIE81,weinberg}.  
In a different context, as mentioned in the introduction, the
consideration of such ``higher'' symmetries  in nuclear structure
physics has provided a remarkable unification of the  spectroscopic
properties of quartets of neighboring nuclei 
\cite{FI}, as we shall explain in the subsequent sections of these 
lecture notes. With this in mind, we shall consider the effects  on the
$\rU^\rB(n)\otimes\rU^\rF(m)$ model  arising from embedding its
dynamical algebra into a superalgebra. 

To start our discussion of superalgebras, it is convenient to consider 
a schematic example, consisting of system formed by 
a single boson and a single (``spinless'') fermion, 
denoted by $\db$ and $\da$, respectively. 
In this case the bilinear products 
$\cG^\rB_\rB=\db b$ and $\cG^\rF_\rF=\da a$ 
each generate a U(1) algebra. 
Taken together, these generators conform the 
\beqn 
\rU^\rB(1)\otimes\rU^\rF(1) ~,
\eeqn 
dynamical algebra, in analogy with the boson--fermion algebra 
mentioned above \cite{nueve}. 
Let us now consider the introduction 
of the mixed terms $\db a$ and $\da b$. 
Computing the commutator of these operators, we find 
\beqn 
[\da b,\db a] 
=\da b\db a-\db a\da b 
=\da a-\db b+2\db b\da a ~,
\eeqn 
which does not close into the original set 
$(\da a,\db b,\da b,\db a)$. 
This means that the inclusion of the cross terms 
does not lead to a Lie algebra. 
We note, however, that the bilinear operators $\db a$ and $\da b$ 
do not behave like bosons, but rather as fermion operators, 
in contrast to $\da a$ and $\db b$, 
both of which have bosonic character 
(in the sense that, e.g., $\da_i a_j$ {\em commutes} with $\da_k a_l)$. 
This suggests the separation of the generators in two {\it sectors}, 
the bosonic sector $(\da a,\db b)$ 
and the fermionic sector $(\da b,\db a)$. 
Computing the {\em anti}commutators of the latter, we find 
 \beq 
\{\da b,\da b\}=0~, 
\quad 
\{\db a,\db a\}=0~, 
\quad 
\{\da b,\db a\} 
=\da a+\db b ~,
\label{4sa3} 
\eeq 
which indeed close into the same set. 
The commutators between the bosonic and fermionic sectors give 
\beq 
\begin{array}{lcl} 
[\da b,\da a]=-\da b~, 
&\qquad& 
[\db a,\da a]=\db a ~,
\\[1ex] 
[\da b,\db b]=\da b~, 
&\qquad& 
[\db a,\db b]=-\db a~. 
\end{array} 
\label{4sa4} 
\eeq 
The operations defined in (\ref{4sa3}) and (\ref{4sa4}), 
together with the (in this case) 
trivial $\rU^\rB(1)\otimes\rU^\rF(1)$ commutators 
\beqn 
[\da a,\da a]=[\db b,\db b]=[\da a,\db b]=0 ~,
\eeqn 
define the {\it superalgebra} U(1/1). 
To maintain the closure property for the enlarged set of generators 
belonging to the boson and fermion sectors, 
we are thus forced to include both commutators and anticommutators 
in the definition of a superalgebra. 
In general, superalgebras then involve boson-sector generators $\hB_i$ 
and fermion-sector generators $\hF_j$, 
satisfying the generalized relations 
\beq 
\displaystyle 
[\hB_i,\hB_j]=\sum_k c^k_{ij}\hB_k~, 
\quad 
[\hB_i,\hF_j]=\sum_k d^k_{ij}\hF_k~, 
\quad 
\{\hF_i,\hF_j\}=\sum_k e^k_{ij}\hB_k~, 
\label{4sa6} 
\eeq 
where $c^k_{ij}$, $d^k_{ij}$, and $e^k_{ij}$ 
are complex constants defining the structure of the superalgebra, 
hence their denomination 
as {\it structure constants} of the superalgebra \cite{4BAR83}. 
We shall only be concerned in these lecture notes 
with superalgebras of the form $\rU(n/m)$, 
where $n$ and $m$ denote the dimensions 
of the boson and fermion subalgebras $\rU^\rB(n)$ and $\rU^\rF(m)$. 
In Section 3 of these notes, we focus our attention on nuclear 
supersymmetry. 

\section{Nuclear Supersymmetry}

Nuclear supersymmetry (n-SUSY) is a composite-particle phenomenon,
linking the properties of bosonic and fermionic systems, framed in
the context of the Interacting Boson Model of nuclear structure \cite{IBM}. 
Composite particles, such as the $\alpha$-particle are  known to
behave as approximate bosons.  As He atoms they become superfluid
at low temperatures, an under certain conditions can  also form
Bose-Einstein condensates.  At higher densities (or temperatures)
the constituent fermions begin to be felt and the  Pauli principle
sets in. Odd-particle composite systems, on the  other hand,
behave as approximate fermions, which in the case of the Interacting
Boson-Fermion Model are treated as a combination of bosons and an 
(ideal) fermion \cite{IBFM}. In
contrast to the theoretical construct of supersymmetric particle
physics, where SUSY is postulated as a generalization of the
Lorentz-Poincare invariance at a fundamental level, experimental
evidence has been found for n-SUSY 
\cite{FI,susy,baha,thesis,tres,pt195,au196} 
as we shall discuss below. 
Nuclear supersymmetry should not be confused with fundamental SUSY, which
predicts the existence of supersymmetric particles, such as the
photino and the selectron for which, up to now, no evidence
has been found. If such particles exist, however, SUSY must be
strongly broken, since large mass differences must exist among
superpartners, or otherwise they would have been already detected.
Competing SUSY models give rise to diverse mass predictions and
are the basis for current superstring and brane
theories \cite{weinberg,cuatro}. Nuclear supersymmetry, on the other hand,
is a theory that establishes precise links among the spectroscopic
properties of certain neighboring nuclei. Even-even and odd-odd
nuclei are composite bosonic systems, while odd-A nuclei are
fermionic. It is in this context that n-SUSY provides a
theoretical framework where bosonic and fermionic systems
are treated as members of the same supermultiplet \cite{baha}.
Nuclear supersymmetry treats the excitation spectra and
transition intensities of the  different nuclei as arising from a
single Hamiltonian and a single set of transition operators. As we 
mentioned before, nuclear SUSY was originally postulated as a
symmetry among pairs of nuclei \cite{FI,susy,baha}, and was
subsequently extended to nuclear quartets or ``magic squares'',
where odd-odd nuclei could be incorporated in a natural
way \cite{quartet}.  Evidence for the existence of n-SUSY (albeit
possibly significantly broken) grew over the years, specially for
the quartet provided by the nuclei $^{194}$Pt, $^{195}$Au, $^{195}$Pt
and $^{196}$Au, but only recently  more systematic evidence
was found.  This was achieved by means of one-nucleon transfer
reaction experiments leading to the odd-odd nucleus $^{196}$Au,
which, together with the other members of the SUSY quartet
 is considered to be the
best example of n-SUSY in nature \cite{quartet,pt195,au196}. We
should point out, however, that while these experiments provided
the first complete energy classification for $^{196}$Au (which was
found to be consistent with the theoretical
predictions \cite{quartet,pt195,au196}), the reactions involved
($^{197}$Au$(\vec d,t)$, $^{197}$Au$(p,d)$ and $^{198}$Hg$(\vec
d,\alpha)$) did not actually test directly the supersymmetric wave
functions. Furthermore, whereas these
new measurements are very exciting, the dynamical SUSY framework
is so restrictive that there was little hope that other quartets
could be found and used to verify the
theory \cite{quartet,pt195,au196}. In the following sections we
emphasize two aspects of SUSY research. On the one hand we
report on an ongoing investigation of one- and two-nucleon
transfer reactions \cite{diez} in the Pt-Au region that will more
 directly analyze the supersymmetric wave functions and
measure new correlations which have not been tested up to now.  On
the other hand we discuss some ideas put forward several years
ago, which question the need for dynamical symmetries in order for
n-SUSY to exist \cite{once,diecinueve}. We thus propose a more general
theoretical framework for nuclear supersymmetry.  The combination
of such a generalized form of supersymmetry and the  transfer
experiments now being carried out \cite{graw}, could provide
remarkable new correlations and a unifying theme in  nuclear
structure physics.

We first present a pedagogic review of dynamical (super)symmetries in 
even- and odd-mass nuclei, which is based in part on \cite{thesis}. 
Next we discuss some new results on correlations between different transfer 
reactions and some perspectives for future work. 

\subsection{Dynamical Symmetries in Even-Even Nuclei}

Dynamical supersymmetries were introduced \cite{FI} in nuclear physics in
1980 by Franco Iachello in the context of the Interacting Boson Model (IBM)
\cite{IBM} and its extensions. The spectroscopy of atomic nuclei is
characterized by the interplay between collective (bosonic) and
single-particle (fermionic) degrees of freedom.

The IBM describes collective excitations in even-even nuclei in
terms of a system of interacting monopole and quadrupole bosons with angular
momentum $l=0,2$. The bosons are associated with the number of
correlated proton and neutron pairs, and hence the number of bosons $N$ is
half the number of valence nucleons. Since it is convenient to express
the Hamiltonian and other operators of interest in second quantized form,
we introduce creation, $s^{\dagger}$ and $d^{\dagger}_m$, and annihilation,
$s$ and $d_m$, operators, which altogether can be denoted by
$b^{\dagger}_{i}$ and $b_{i}$ with $i=l,m$ ($l=0,2$ and $-l \leq m \leq l$).
The operators $b^{\dagger}_{i}$ and $b_{i}$ satisfy the commutation
relations
\begin{equation}
[b_i,b^{\dagger}_j] \;=\; \delta_{ij} ~,
\hspace{1cm} [b^{\dagger}_i,b^{\dagger}_j]
\;=\; [b_i,b_j] \;=\; 0 ~.
\end{equation}
The bilinear products
\begin{equation}
B_{ij} \;=\; b^{\dagger}_i b_j ~,
\label{bosgen}
\end{equation}
generate the algebra of $U(6)$ the unitary group in 6 dimensions
\begin{equation}
[ B_{ij},B_{kl} ] \;=\; B_{il} \, \delta_{jk} - B_{kj} \, \delta_{il} ~.
\end{equation}
We want to construct states and operators that transform according
to irreducible representations of the rotation group (since the
problem is rotationally invariant).
The creation operators $b^{\dagger}_i$ transform by definition as
irreducible tensors under rotation. However, the annihilation operators
$b_i$ do not. It is an easy exercise to contruct operators that do
transform appropriately
\begin{equation}
\tilde{b}_{lm} \;=\; (-)^{l-m} b_{l,-m} ~.
\end{equation}
The 36 generators of Eq.~(\ref{bosgen}) can be rewritten in
angular-momentum-coupled form as
\begin{equation}
[ b^{\dagger}_{l} \times \tilde{b}_{l'} ]^{(L)}_{M} \;=\;
\sum_{mm'} \langle l,m,l',m' | L,M \rangle \,
b^{\dagger}_{lm} \tilde{b}_{l'm'} ~.
\end{equation}
The one- and two-body Hamiltonian can be expressed in terms
of the generators of $U(6)$ as
\begin{eqnarray}
H &=& \sum_l \epsilon_l \sum_m b^{\dagger}_{lm} b_{lm}
\nonumber\\
&& + \sum_L \sum_{l_1 l_2 l_3 l_4} u^{(L)}_{l_1 l_2 l_3 l_4} \,
[ [b^{\dagger}_{l_1} \times \tilde{b}_{l_2} ]^{(L)} \times
[b^{\dagger}_{l_3} \times \tilde{b}_{l_4} ]^{(L)} ]^{(0)} ~.
\label{hb}
\end{eqnarray}
In general, the Hamiltonian has to be diagonalized numerically to
obtain the energy eigenvalues and wave functions. There exist, however,
special situations in which the eigenvalues can be obtained in closed,
analytic form. These special solutions provide a framework in which
energy spectra and other nuclear properties (such as quadrupole transitions
and moments) can be interpreted in a qualitative way.
These situations correspond to dynamical symmetries of the Hamiltonian
\cite{IBM} (see section~\ref{algebraic}). 

The concept of dynamical symmetry has been shown to be a very useful tool
in different branches of physics. A well-known example in nuclear physics
is the Elliott $SU(3)$ model \cite{Elliott} to describe the properties
of light nuclei in the $sd$ shell. Another example is the $SU(3)$ flavor
symmetry of Gell-Mann and Ne'eman \cite{aGEL64} to classify the baryons
and mesons into flavor octets, decuplets and singlets and to describe
their masses with the Gell-Mann-Okubo mass formula, as described in the 
previous sections. 

The group structure
of the IBM Hamiltonian is that of $G=U(6)$. Since nuclear states have good
angular momentum, the rotation group in three dimensions $SO(3)$ should be
included in all subgroup chains of $G$ \cite{IBM}
\begin{eqnarray}
U(6)  \supset \left\{ \begin{array}{l}
U(5)  \supset SO(5) \supset SO(3) ~,\\
SO(6) \supset SO(5) \supset SO(3) ~,\\
SU(3) \supset SO(3) ~.
\end{array} \right.
\label{bchains}
\end{eqnarray}
The three dynamical symmetries which correspond to the group chains in
Eq.~(\ref{bchains}) are limiting cases of the IBM and are usually
referred to as the $U(5)$ (vibrator), the $SU(3)$ (axially symmetric rotor)
and the $SO(6)$ ($\gamma$-unstable rotor).

Here we consider a simplified form of the general expression of
the IBM Hamiltonian of Eq.~(\ref{hb}) that contains the main
features of collective motion in nuclei
\begin{equation}
H \;=\; \epsilon \, \hat n_d - \kappa \, \hat Q(\chi) \cdot \hat Q(\chi) ~,
\label{cqm}
\end{equation}
where $n_d$ counts the number of quadrupole bosons
\begin{equation}
\hat n_d \;=\; \sqrt{5} \, [d^{\dagger} \times \tilde{d} ]^{(0)}
\;=\; \sum_m d^{\dagger}_m d_m ~,
\end{equation}
and $Q$ is the quadrupole operator
\begin{equation}
\hat Q_m(\chi) \;=\; [ s^{\dagger} \times \tilde{d}
+ d^{\dagger} \times \tilde{s}
+ \chi \, d^{\dagger} \times \tilde{d} ]^{(2)}_m ~.
\end{equation}
The three dynamical symmetries are recovered for different choices
of the coefficients $\epsilon$, $\kappa$ and $\chi$. Since the IBM
Hamiltonian conserves the number of bosons and is invariant under
rotations, its eigenstates can be labeled by the total number of
bosons $N$ and the angular momentum $L$.

In the absence of a quadrupole-quadrupole interaction $\kappa=0$,
the Hamiltonian of Eq.~(\ref{cqm}) becomes proportional to the
linear Casimir operator of $U(5)$
\begin{equation}
H_1 \;=\; \epsilon \, \hat n_d \;=\; \epsilon \, {\cal C}_{1U(5)} ~.
\end{equation}
In addition to $N$, $L$ and $M$, the basis states can be labeled by
the quantum numbers $n_d$ and $\tau$, which characterize the irreducible
representations of $U(5)$ and $SO(5)$. Here $n_d$ represents the number
of quadrupole bosons and $\tau$ the boson seniority. The eigenvalues
of $H_1$ are given by the expectation value of the Casimir operator
\begin{equation}
E_1 \;=\; \epsilon \, n_d ~.
\end{equation}
In this case, the energy spectrum is characterized by a series of
multiplets, labeled by the number of quadrupole bosons, at a constant
energy spacing which is typical for a vibrational nucleus.

For the quadrupole-quadrupole interaction, we can distinguish two
situations in which the eigenvalue problem can be solved analytically.
If $\chi=\mp \sqrt{7}/2$, the Hamiltonian has a $SU(3)$ dynamical symmetry
\begin{equation}
H_2 \;=\; - \kappa \, \hat Q(\mp \sqrt{7}/2) \cdot
\hat Q(\mp \sqrt{7}/2) \;=\; -\frac{1}{2} \kappa
\left[ {\cal C}_{2SU(3)} - \frac{3}{4} {\cal C}_{2SO(3)} \right] ~.
\end{equation}
In this case, the eigenstates can be labeled by $(\lambda,\mu)$
which characterize the irreducible representations of $SU(3)$.
The eigenvalues are
\begin{equation}
E_2 \;=\; -\frac{1}{2} \kappa \left[
\lambda(\lambda+3)+\mu(\mu+3)+\lambda\mu)
-\frac{3}{4} \kappa L(L+1) \right] ~.
\end{equation}
The energy spectrum is characterized by a series of bands, in which
the energy spacing is proportional to $L(L+1)$, as in the rigid rotor
model. The ground state band has $(\lambda,\mu)=(2N,0)$ and the first
excited band $(2N-4,2)$ corresponds to a degenerate $\beta$ and $\gamma$
band. The sign of the coefficient $\chi$ is related to a prolate (-) or
an oblate (+) deformation.

For $\chi=0$, the Hamiltonian has a $SO(6)$ dynamical symmetry
\begin{equation}
H_3 \;=\; -\kappa \, \hat Q(0) \cdot \hat Q(0)
\;=\; -\kappa \left[ {\cal C}_{2SO(6)} - {\cal C}_{2SO(5)} \right] ~.
\end{equation}
The basis states are labeled by $\sigma$ and $\tau$
which characterize the irreducible representations of $SO(6)$ and $SO(5)$,
respectively. Characteristic features of the energy spectrum
\begin{equation}
E_3 \;=\; -\kappa \left[ \sigma(\sigma+4)-\tau(\tau+3) \right] ~,
\end{equation}
are the repeating patterns $L=0,2,4,2$ which is typical of the
$\gamma$-unstable rotor.

For other choices of the coefficients, the Hamiltonian of Eq.~(\ref{cqm})
describes situations in between any of the dynamical symmetries which
correspond to transitional regions, e.g. the Pt-Os isotopes exhibit a
transition between a $\gamma$-unstable and a rigid rotor
$SO(6) \leftrightarrow SU(3)$, the Sm isotopes between vibrational and
rotational nuclei $U(5) \leftrightarrow SU(3)$, and the Ru isotopes
between vibrational and $\gamma$-unstable nuclei
$U(5) \leftrightarrow SO(6)$.

\subsection{Dynamical Symmetries in Odd-A Nuclei}

For odd-mass nuclei the IBM has been extended to include single-particle
degrees of freedom \cite{IBFM}. The Interacting Boson-Fermion Model (IBFM)
has as its building blocks a set of $N$ bosons with $l=0,2$ and an odd
nucleon (either a proton or a neutron) occupuying the single-particle
orbits with angular momenta $j=j_1,j_2,\dots$. The components of the
fermion angular momenta span the $m$-dimensional space of the group
$U(m)$ with $m=\sum_j (2j+1)$.

We introduce, in addition to the boson creation $b^{\dagger}_i$ and
annihilation $b_i$ operators for the collective degrees of freedom,
fermion creation $a^{\dagger}_i$ and annihilation $a_{i}$ operators
for the single-particle. The fermion operators satisfy
anti-commutation relations
\begin{equation}
\{a_i,a^{\dagger}_j\} \;=\; \delta_{ij} ~,
\hspace{1cm} \{a^{\dagger}_i,a^{\dagger}_j\}
\;=\; \{a_i,a_j\} \;=\; 0 ~.
\end{equation}
By construction the fermion operators commute with the boson operators.
The bilinear products
\begin{equation}
A_{ij} \;=\; a^{\dagger}_i a_j ~,
\label{fergen}
\end{equation}
generate the algebra of $U(m)$, the unitary group in $m$ dimensions
\begin{equation}
[ A_{ij},A_{kl} ] \;=\; A_{il} \, \delta_{jk} - A_{kj} \, \delta_{il} ~.
\end{equation}
For the mixed system of boson and fermion degrees of freedom we introduce
angular-momentum-coupled generators as
\begin{equation}
B^{(L)}_M(l,l') \;=\; [ b^{\dagger}_{l} \times \tilde{b}_{l'} ]^{(L)}_{M} ~,
\hspace{1cm}
A^{(L)}_M(j,j') \;=\; [ a^{\dagger}_{j} \times \tilde{a}_{j'} ]^{(L)}_{M} ~,
\end{equation}
where $\tilde{a}_{jm}$ is defined to be a spherical tensor operator
\begin{equation}
\tilde{a}_{jm} \;=\; (-)^{j-m} a_{j,-m} ~.
\end{equation}
The most general one- and two-body rotational invariant Hamiltonian
of the IBFM can be written as
\begin{equation}
H \;=\; H_B + H_F + V_{BF} ~,
\end{equation}
where $H_B$ is the IBM Hamiltonian of Eq.~(\ref{hb}), $H_F$ is the
fermion Hamiltonian
\begin{eqnarray}
H_F &=& \sum_j \eta_j \sum_m a^{\dagger}_{jm} a_{jm}
\nonumber\\
&& + \sum_L \sum_{j_1 j_2 j_3 j_4} v^{(L)}_{j_1 j_2 j_3 j_4} \,
[ [a^{\dagger}_{j_1} \times \tilde{a}_{j_2} ]^{(L)} \times
[a^{\dagger}_{j_3} \times \tilde{a}_{j_4} ]^{(L)} ]^{(0)} ~,
\label{hf}
\end{eqnarray}
and $B_{BF}$ the boson-fermion interaction
\begin{eqnarray}
V_{BF} &=& \sum_L \sum_{l_1 l_2 j_1 j_2} w^{(L)}_{l_1 l_2 j_1 j_2} \,
[ [b^{\dagger}_{l_1} \times \tilde{b}_{l_2} ]^{(L)} \times
[a^{\dagger}_{j_1} \times \tilde{a}_{j_2} ]^{(L)} ]^{(0)} ~.
\label{vbf}
\end{eqnarray}

The IBFM Hamiltonian has an interesting algebraic structure, that suggests
the possible occurrence of dynamical symmetries in odd-A nuclei.
Since in the IBFM odd-A nuclei are described in terms of a mixed system
of interacting bosons and fermions, the concept of dynamical symmetries
has to be generalized. Under the restriction, that both the boson and
fermion states have good angular momentum, the respective group chains
should contain the rotation group ($SO(3)$ for bosons and $SU(2)$ for
fermions) as a subgroup
\begin{eqnarray}
U^B(6) \supset \ldots \supset SO^B(3) ~,
\nonumber\\
U^F(m) \supset \ldots \supset SU^F(2) ~,
\end{eqnarray}
where we have introduced superscripts to  distinguish between boson
and fermion groups.
If one of subgroups of $U^B(6)$ is isomorphic
to one of the subgroups of $U^F(m)$, the boson and fermion group chains
can be combined into a common boson-fermion group chain. When the
Hamiltonian is written in terms of Casimir invariants of the combined
boson-fermion group chain, a dynamical boson-fermion symmetry arises.

Among the many different possibilities, we consider two dynamical
boson-fermion symmetries associated with the $SO(6)$ limit of the IBM.
The first example discussed in the literature \cite{FI,spin6} is the case
of bosons with $SO(6)$ symmetry and the odd nucleon occupying a
single-particle orbit with spin $j=3/2$. The relevant group chains are
\begin{eqnarray}
U^B(6) &\supset& SO^B(6) \supset SO^B(5) \supset SO^B(3) ~,
\nonumber\\
U^F(4) &\supset& SU^F(4) \supset Sp^F(4) \supset SU^F(2) ~.
\end{eqnarray}
Since $SO(6)$ and $SU(4)$ are isomorphic, the boson and fermion group
chains can be combined into
\begin{eqnarray}
U^{B}(6) \otimes U^{F}(4) &\supset& SO^{B}(6) \otimes SU^{F}(4)
\nonumber\\
&\supset& Spin(6) \supset Spin(5) \supset Spin(3) ~.
\end{eqnarray}
The spinor groups $Spin(n)$ are the universal covering groups of the
orthogonal groups $SO(n)$, with $Spin(6) \sim SU(4)$, $Spin(5) \sim Sp(4)$
and $Spin(3) \sim SU(2)$. The generators of the
spinor groups consist of the sum of a boson and a fermion part.
For example, for the quadrupole operator we have
\begin{equation}
\hat Q_m \;=\; [ s^{\dagger} \times \tilde{d}
+ d^{\dagger} \times \tilde{s} ]^{(2)}_m
+ [ a^{\dagger}_{3/2} \times \tilde{a}_{3/2} ]^{(2)}_m ~.
\end{equation}
We consider a simple quadrupole-quadrupole interaction which, just as for
the $SO(6)$ limit of the IBM, can be written as the difference of two
Casimir invariants
\begin{equation}
H \;=\; -\kappa \, \hat Q \cdot \hat Q
\;=\; -\kappa \left[ {\cal C}_{2Spin(6)} - {\cal C}_{2Spin(5)} \right] ~.
\end{equation}
The basis states are classified by $(\sigma_1,\sigma_2,\sigma_3)$,
$(\tau_1,\tau_2)$ and $J$ which label the irreducible representations
of the spinor groups $Spin(6)$, $Spin(5)$ and $Spin(3)$.
The energy spectrum is obtained from the expectation value of the
Casimir invariants of the spinor groups
\begin{equation}
E \;=\; -\kappa \left[ \sigma_1(\sigma_1+4) + \sigma_2(\sigma_2+2)
+ \sigma_3^2 - \tau_1(\tau_1+3) - \tau_2(\tau_2+1) \right] ~.
\end{equation}
The mass region of the Os-Ir-Pt-Au nuclei, where the even-even Pt nuclei
are well described by the $SO(6)$ limit of the IBM and the odd proton
mainly occupies the $d_{3/2}$ shell, seems to provide
experimental examples of this symmetry, e.g. $^{191,193}$Ir and
$^{193,195}$Au.

The concept of dynamical boson-fermion symmetries is not restricted to
cases in which the odd nucleon occupies only a single-$j$ orbit. The
first example of a multi-$j$ case discussed in the literature \cite{baha}
is that of a dynamical boson-fermion symmetry associated with the $SO(6)$
limit and the odd nucleon occupying single-particle orbits with spin
$j=1/2$, 3/2, 5/2. In this case, the fermion space is decomposed into a
pseudo-orbital part with $k=0,2$ and a pseudo-spin part with $s=1/2$
corresponding to the group reduction
\begin{eqnarray}
U^F(12) \supset U^F(6) \otimes U^F(2) \supset \left\{
\begin{array}{c} U^F(5) \otimes U^F(2) ~, \\
SU^F(3) \otimes U^F(2) ~, \\ SO^F(6) \otimes U^F(2) ~.
\end{array} \right.
\end{eqnarray}
Since the pseudo-orbital angular momentum $k$ has the same values as
the angular momentum of the $s$- and $d$- bosons of the IBM, it is clear
that the pseudo-orbital part can be combined with all three dynamical
symmetries of the IBM.
\begin{eqnarray}
U^B(6) \supset \left\{ \begin{array}{c} U^B(5) ~, \\
SU^B(3) ~, \\ SO^B(6) ~. \end{array} \right.
\end{eqnarray}
into a dynamical boson-fermion symmetry. The case, in which the bosons
have $SO(6)$ symmetry is of particular interest, since the negative
parity states in Pt with the odd neutron occupying the $3p_{1/2}$,
$3p_{3/2}$ and $3f_{5/2}$ orbits have been suggested as possible
experimental examples of a multi-$j$ boson-fermion symmetry.
In this case, the relevant boson-fermion group chain is
\begin{eqnarray}
U^{B}(6) \otimes U^{F}(12)
&\supset& U^B(6) \otimes U^{F}(6) \otimes U^{F}(2)
\nonumber\\
&\supset& U^{BF}(6) \otimes U^{F}(2)
\nonumber\\
&\supset& SO^{BF}(6) \otimes U^{F}(2)
\nonumber\\
&\supset& SO^{BF}(5) \otimes U^{F}(2)
\nonumber\\
&\supset& SO^{BF}(3) \otimes SU^{F}(2)
\nonumber\\
&\supset& SU(2) ~.
\end{eqnarray}
Just as in the first example for the spinor groups, the generators of
the boson-fermion groups consist of the sum of a boson and a fermion part,
e.g. the quadrupole operator is now written as
\begin{eqnarray}
\hat Q_m \;=\; [ s^{\dagger} \times \tilde{d}
+ d^{\dagger} \times \tilde{s} ]^{(2)}_m
&+& \sqrt{\frac{4}{5}} \, [ a^{\dagger}_{3/2} \times \tilde{a}_{1/2}
- a^{\dagger}_{1/2} \times \tilde{a}_{3/2} ]^{(2)}_m
\nonumber\\
&+& \sqrt{\frac{6}{5}} \, [ a^{\dagger}_{5/2} \times \tilde{a}_{1/2}
+ a^{\dagger}_{1/2} \times \tilde{a}_{5/2} ]^{(2)}_m ~.
\end{eqnarray}
Also in this case, the quadrupole-quadrupole interaction can be written
as the difference of two Casimir invariants
\begin{equation}
H \;=\; -\kappa \, \hat Q \cdot \hat Q
\;=\; -\kappa \left[ {\cal C}_{2SO^{BF}(6)} - {\cal C}_{2SO^{BF}(5)} \right] ~.
\end{equation}
The basis states are classified by $(\sigma_1,\sigma_2,\sigma_3)$,
$(\tau_1,\tau_2)$ and $L$ which label the irreducible representations
of the boson-fermion groups $SO^{BF}(6)$, $SO^{BF}(5)$ and $SO^{BF}(3)$.
Although the labels are the same as for the previous case, the allowed
values are different.
The total angular momentum is given by $\vec{J}=\vec{L}+\vec{s}$.
The energy spectrum is given by
\begin{equation}
E \;=\; -\kappa \left[ \sigma_1(\sigma_1+4) + \sigma_2(\sigma_2+2)
+ \sigma_3^2 - \tau_1(\tau_1+3) - \tau_2(\tau_2+1) \right] ~.
\end{equation}
The mass region of the Os-Ir-Pt-Au nuclei, where the even-even Pt nuclei
are well described by the $SO(6)$ limit of the IBM and the odd neutron
mainly occupies the negative parity orbits $3p_{1/2}$, $3p_{3/2}$ and
$3f_{5/2}$ provides experimental examples of this symmetry, in particular
the nucleus $^{195}$Pt \cite{baha,BI,sun,pt195}

\subsection{Dynamical Supersymmetries}

Boson-fermion symmetries can further be extended by introducing the concept
of supersymmetries \cite{susy}, in which states in both even-even and
odd-even nuclei are treated in a single framework. In the previous section,
we have discussed the symmetry properties of a mixed system of boson and
fermion degrees of freedom for a fixed number of bosons $N$ and one fermion
$M=1$. The operators $B_{ij}$ and $A_{ij}$
\begin{equation}
B_{ij} \;=\; b^{\dagger}_i b_j ~, \hspace{1cm}
A_{ij} \;=\; a^{\dagger}_i a_j ~,
\label{bgen}
\end{equation}
which generate the Lie algebra of the symmetry group $U^B(6) \otimes U^F(m)$
of the IBFM, can only change
bosons into bosons and fermions into fermions. The number of bosons $N$ and
the number of fermions $M$ are both conserved quantities. As explained in 
Section 2.6, in addition to $B_{ij}$ and $A_{ij}$, one can introduce 
operators that change a boson into a fermion and vice versa
\begin{equation}
F_{ij} \;=\; b^{\dagger}_i a_j ~, \hspace{1cm}
G_{ij} \;=\; a^{\dagger}_i b_j ~.
\label{fgen}
\end{equation}
The enlarged set of operators $B_{ij}$, $A_{ij}$, $F_{ij}$ and $G_{ij}$
forms a closed algebra which consists of both commutation and
anticommutation relations
\begin{eqnarray}
\, [ B_{ij}, B_{kl} ] &=& B_{il} \, \delta_{jk} - B_{kj} \, \delta_{il} ~,
\nonumber\\
\, [ B_{ij}, A_{kl} ] &=& 0 ~,
\nonumber\\
\, [ B_{ij}, F_{kl} ] &=& F_{il} \, \delta_{jk} ~,
\nonumber\\
\, [ B_{ij}, G_{kl} ] &=& -G_{kj} \, \delta_{il} ~,
\nonumber\\
\, [ A_{ij}, A_{kl} ] &=& A_{il} \, \delta_{jk} - A_{kj} \, \delta_{il} ~,
\nonumber\\
\, [ A_{ij}, F_{kl} ] &=& -F_{kj} \, \delta_{il} ~,
\nonumber\\
\, [ A_{ij}, G_{kl} ] &=& G_{il} \, \delta_{jk} ~,
\nonumber\\
\{ F_{ij}, F_{kl} \}  &=& 0 ~,
\nonumber\\
\{ F_{ij}, G_{kl} \}  &=& B_{il} \, \delta_{jk} + A_{kj} \, \delta_{il} ~,
\nonumber\\
\{ G_{ij}, G_{kl} \}  &=& 0 ~.
\end{eqnarray}
This algebra can be identified with that of the graded Lie group $U(6/m)$.
It provides an elegant scheme in which the IBM and IBFM can be unified
into a single framework \cite{susy}
\begin{equation}
U(6/m) \supset U^B(6) \otimes U^F(m) ~.
\end{equation}
In this supersymmetric framework,
even-even and odd-mass nuclei form the members of a supermultiplet which
is characterized by ${\cal N}=N+M$, i.e. the total number of bosons and
fermions. Supersymmetry thus distinguishes itself from ``normal'' symmetries
in that it includes, in addition to transformations among fermions and among
bosons, also transformations that change a boson into a fermion and
vice versa.

The Os-Ir-Pt-Au mass region provides ample
experimental evidence for the occurrence of dynamical (super)symmetries in
nuclei. The even-even nuclei $^{194,196}$Pt are the standard examples of
the $SO(6)$ limit of the IBM \cite{so6} and the odd proton, in first
approximation, occupies the single-particle level $2d_{3/2}$. In this
special case, the boson and fermion groups can be combined into spinor groups,
and the odd-proton nuclei $^{191,193}$Ir and $^{193,195}$Au were suggested
as examples of the $Spin(6)$ limit \cite{FI}. The appropriate extension
to a supersymmetry is by means of the graded Lie group $U(6/4)$
\begin{eqnarray}
U(6/4) \supset U^{B}(6) \otimes U^{F}(4)
&\supset& SO^{B}(6) \otimes SU^{F}(4)
\nonumber\\
&\supset& Spin(6) \supset Spin(5) \supset Spin(3) \supset Spin(2) ~.
\end{eqnarray}
The pairs of nuclei $^{190}$Os - $^{191}$Ir, $^{192}$Os - $^{193}$Ir,
$^{192}$Pt - $^{193}$Au and $^{194}$Pt - $^{195}$Au have been analyzed as
examples of a $U(6/4)$ supersymmetry \cite{susy}.

\begin{figure}[t]
\centering
$\begin{array}{c}
\begin{turn}{-90}
\includegraphics[height=10cm]{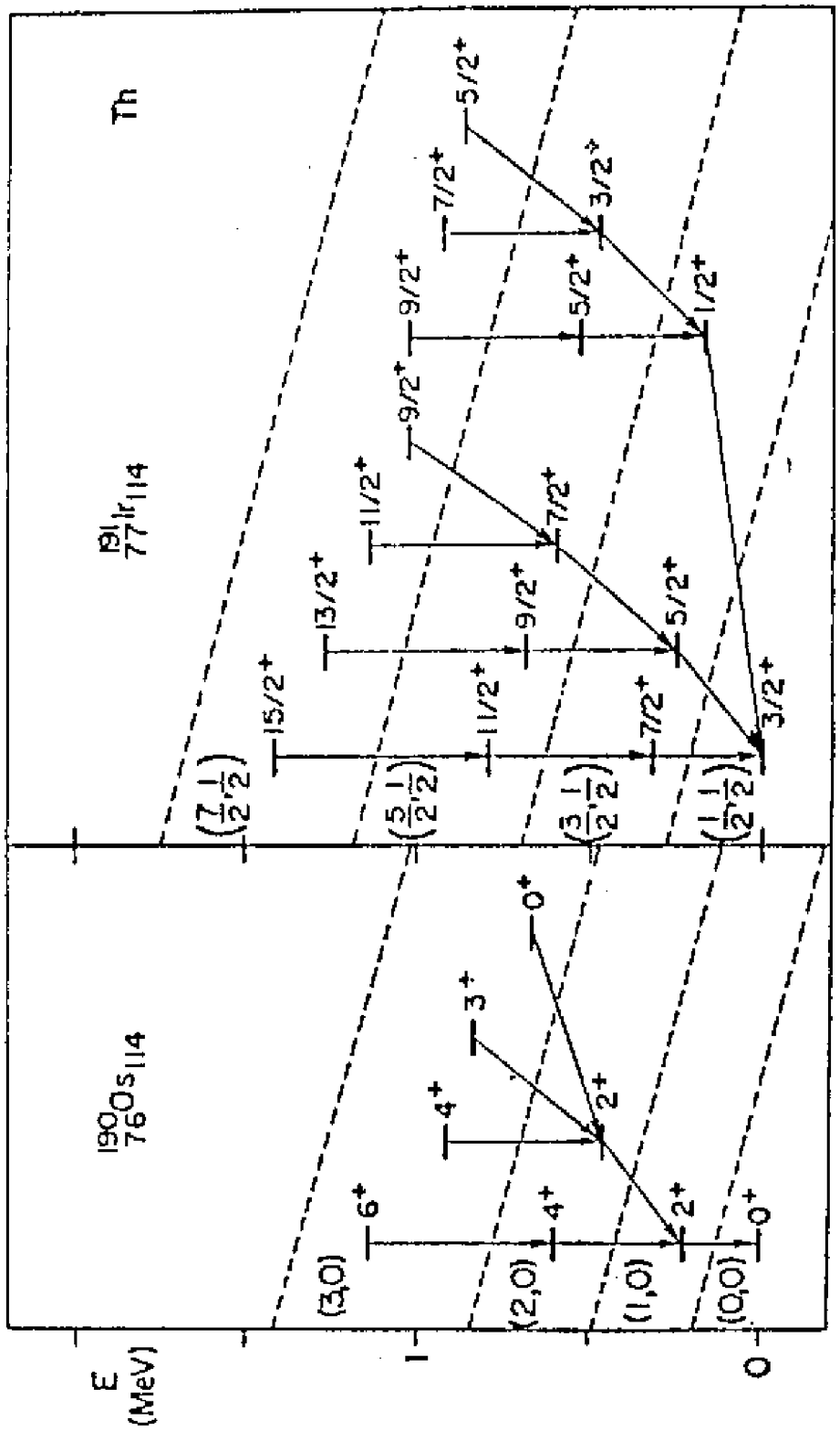}
\end{turn} \\
\hspace{0.6cm}
\begin{turn}{ 90}
\includegraphics[height=10.5cm]{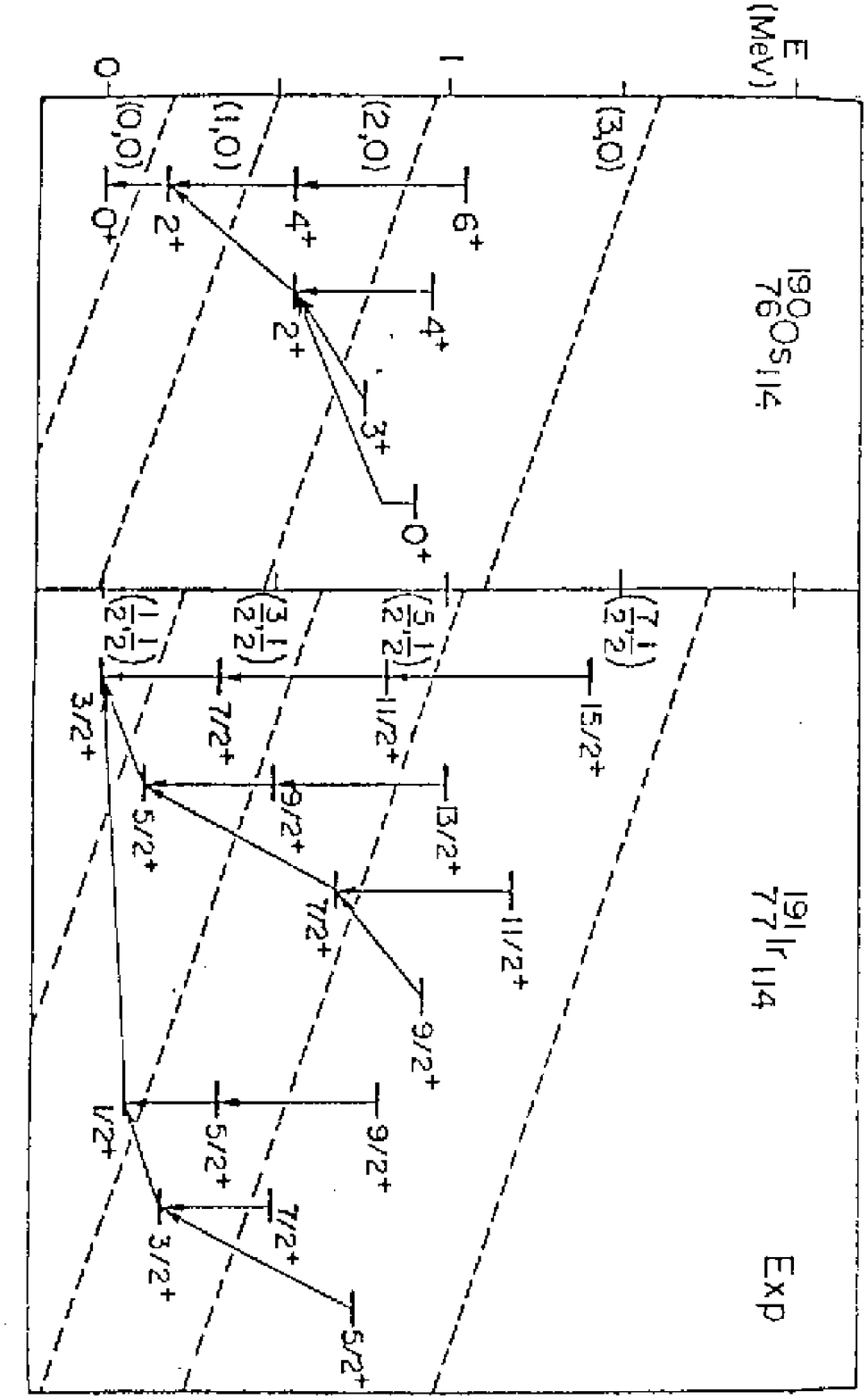}
\end{turn}
\end{array}$
\caption[]{\small Example of a $U(6/4)$ supersymmetry}
\end{figure}

Another example of a dynamical supersymmetry in this mass region
is that of the Pt nuclei. The even-even isotopes are well described
by the $SO(6)$ limit of the IBM and the odd neutron mainly occupies
the negative parity orbits $3p_{1/2}$, $3p_{3/2}$ and $3f_{5/2}$.
In this case, the graded Lie group is $U(6/12)$
\begin{eqnarray}
U(6/12) \supset U^{B}(6) \otimes U^{F}(12)
&\supset& U^B(6) \otimes U^{F}(6) \otimes U^{F}(2)
\nonumber\\
&\supset& U^{BF}(6) \otimes U^{F}(2)
\nonumber\\
&\supset& SO^{BF}(6) \otimes U^{F}(2)
\nonumber\\
&\supset& SO^{BF}(5) \otimes U^{F}(2)
\nonumber\\
&\supset& SO^{BF}(3) \otimes SU^{F}(2)
\nonumber\\
&\supset& SU(2) ~.
\end{eqnarray}
The odd-neutron nucleus $^{195}$Pt, together with $^{194}$Pt, were
studied as an example of a $U(6/12)$ supersymmetry \cite{baha,BI,sun}.

\begin{figure}[t]
\centering
\includegraphics[height=10cm]{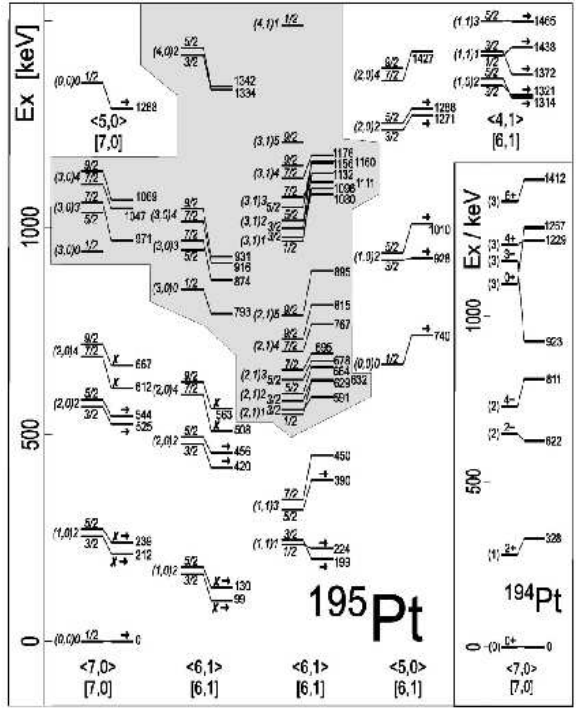}
\caption[]{\small Example of a $U(6/12)$ supersymmetry}
\end{figure}

\subsection{Dynamical Neutron-Proton Supersymmetries}

As we have seen in the previous section, the mass region $A \sim 190$ has
been a rich source of possible empirical evidence for the existence of
(super)symmetries in nuclei. The pairs of nuclei $^{190}$Os - $^{191}$Ir,
$^{192}$Os - $^{193}$Ir, $^{192}$Pt - $^{193}$Au and $^{194}$Pt - $^{195}$Au
have been analyzed as examples of a $U(6/4)$ supersymmetry \cite{susy},
and the nuclei $^{194}$Pt - $^{195}$Pt as an example of a $U(6/12)$
supersymmetry \cite{baha}.
These ideas were later extended to the case where neutron and proton bosons
are distinguished \cite{quartet}, predicting in this way a correlation among
quartets of nuclei, consisting of an even-even, an odd-proton, an odd-neutron
and an odd-odd nucleus. The best experimental example of such a
quartet with $U(6/12)_{\nu} \otimes U(6/4)_{\pi}$ supersymmetry is provided
by the nuclei $^{194}$Pt, $^{195}$Au, $^{195}$Pt and $^{196}$Au. 
\begin{eqnarray}
\begin{array}{ccccc}
\mbox{even-even} & & & & \mbox{odd-even} \\
N_{\nu}+1, N_{\pi}+1 & & & & N_{\nu}, N_{\pi}+1, j_{\nu} \\
& & & & \\
& ^{194}_{ 78}\mbox{Pt}_{116} & \leftrightarrow
& ^{195}_{ 78}\mbox{Pt}_{117} & \\
& & & & \\
& \updownarrow & & \updownarrow & \\
& & & & \\
& ^{195}_{ 79}\mbox{Au}_{116} & \hspace{1cm} \leftrightarrow \hspace{1cm}
& ^{196}_{ 79}\mbox{Au}_{117} & \\
& & & & \\
\mbox{even-odd} & & & & \mbox{odd-odd} \\
N_{\nu}+1, N_{\pi}, j_{\pi} & & & & N_{\nu}, N_{\pi}, j_{\nu}, j_{\pi} 
\end{array}
\label{magic}
\end{eqnarray}
In previous sections, we have used a schematic Hamiltonian consisting
only of a quadrupole-quadrupole interaction to discuss the different
dynamical symmetries. In general, a dynamical (super)symmetry arises
whenever the Hamiltonian is expressed in terms of the Casimir invariants
of the subgroups in a group chain. The relevant subgroup chain of
$U(6/12)_{\nu} \otimes U(6/4)_{\pi}$ for the Pt and Au
nuclei is given by \cite{quartet}
\begin{eqnarray}
U(6/12)_{\nu} \otimes U(6/4)_{\pi} &\supset&
U^{B_{\nu}}(6) \otimes U^{F_{\nu}}(12) \otimes
U^{B_{\pi}}(6) \otimes U^{F_{\pi}}(4)
\nonumber\\
&\supset& U^B(6) \otimes U^{F_{\nu}}(6) \otimes U^{F_{\nu}}(2) \otimes
U^{F_{\pi}}(4)
\nonumber\\
&\supset& U^{BF_{\nu}}(6) \otimes U^{F_{\nu}}(2) \otimes U^{F_{\pi}}(4)
\nonumber\\
&\supset& SO^{BF_{\nu}}(6) \otimes U^{F_{\nu}}(2) \otimes SU^{F_{\pi}}(4)
\nonumber\\
&\supset& Spin(6) \otimes U^{F_{\nu}}(2)
\nonumber\\
&\supset& Spin(5) \otimes U^{F_{\nu}}(2)
\nonumber\\
&\supset& Spin(3) \otimes SU^{F_{\nu}}(2)
\nonumber\\
&\supset& SU(2) ~.
\end{eqnarray}
In this case, the Hamiltonian
\begin{eqnarray}
H &=& \alpha \, C_{2U^{BF_{\nu}}(6)} + \beta \, C_{2SO^{BF_{\nu}}(6)}
+ \gamma \, C_{2Spin(6)}
\nonumber\\
&& + \delta \, C_{2Spin(5)} + \epsilon \, C_{2Spin(3)}
+ \eta \, C_{2SU(2)} ~,
\end{eqnarray}
describes simultaneously the excitation spectra of the quartet of nuclei.
Here we have neglected terms that only contribute to binding energies.
The energy spectrum is given by the eigenvalues of the Casimir operators
\begin{eqnarray}
E &=& \alpha \, \left[ N_1(N_1+5) + N_2(N_2+3) + N_3(N_3+1) \right]
\nonumber\\
&& + \beta \, \left[ \Sigma_1(\Sigma_1+4) + \Sigma_2(\Sigma_2+2)
+ \Sigma_3^2 \right]
\nonumber\\
&& + \gamma \, \left[ \sigma_1(\sigma_1+4) + \sigma_2(\sigma_2+2)
+ \sigma_3^2 \right]
\nonumber\\
&& + \delta \, \left[ \tau_1(\tau_1+3) + \tau_2(\tau_2+1) \right]
+ \epsilon \, J(J+1) + \eta \, L(L+1) ~.
\label{npsusy}
\end{eqnarray}
The coefficients $\alpha$, $\beta$, $\gamma$, $\delta$, $\epsilon$ and
$\eta$ have been determined in a simultaneous fit of the excitation energies
of the four nuclei of Eq.~(\ref{magic}) \cite{au196}.

The supersymmetric classification of nuclear levels in the Pt
and Au isotopes has been re-examined by taking advantage of the significant
improvements in experimental capabilities developed in the last decade.
High resolution transfer experiments with protons and polarized deuterons
have strengthened the evidence for the existence of supersymmetry in
atomic nuclei. The experiments include high resolution transfer experiments
to $^{196}$Au at TU/LMU M\"unchen \cite{tres,pt195}, and in-beam gamma ray
and conversion electron spectroscopy following the reactions
$^{196}$Pt$(d,2n)$ and $^{196}$Pt$(p,n)$ at the cyclotrons of the PSI and
Bonn \cite{au196}. These studies have achieved an improved classification
of states in $^{195}$Pt and $^{196}$Au which give further support to the
original ideas \cite{baha,sun,quartet} and extend and refine previous
experimental work \cite{mauthofer,jolie,rotbard} in this research area.

\begin{figure}[t]
\centering
\includegraphics[height=5cm]{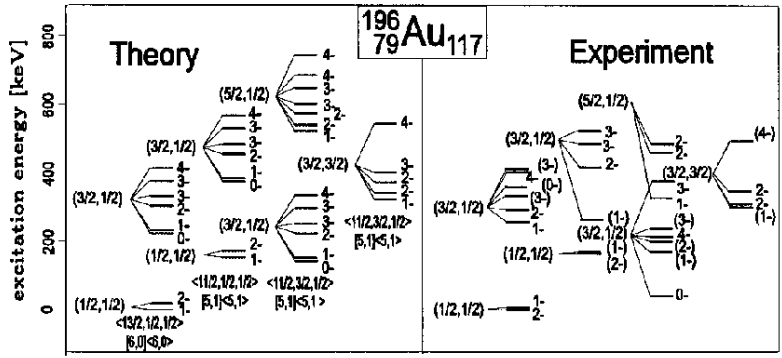}
\caption[]{\small Comparison between the energy spectrum of the negative
parity levels in the odd-odd nucleus $^{196}$Au and that obtained for the
$U(6/12)_{\nu} \otimes U(6/4)_{\pi}$ supersymmetry using Eq.~(\ref{npsusy})
with $\alpha=52.5$, $\beta=8.7$, $\gamma=-53.9$, $\delta=48.8$, 
$\epsilon=8.8$ and $\eta=4.5$ in keV.}
\end{figure}

In analogy to the case of dynamical symmetries, in 
a dynamical supersymmetry closed expressions can be derived for energies, 
as well as selection rules and intensities for electromagnetic transitions and
single-particle transfer reactions. While a simultaneous description and
classification of these observables in terms of the
$U(6/12)_{\nu} \otimes U(6/4)_{\pi}$ supersymmetry has been shown to
be fulfilled to a good approximation for the quartet of nuclei $^{194}$Pt,
$^{195}$Au, $^{195}$Pt and $^{196}$Au, there are important predictions
still not fully verified by experiments. These tests involve the transfer
reaction intensities among the supersymmetric partners. In the next section
we concentrate on the latter and, in particular, on the one-proton transfer
reactions $^{194}$Pt $\rightarrow$ $^{195}$Au and
$^{195}$Pt $\rightarrow$ $^{196}$Au.

\subsection{One-nucleon transfer reactions}

The single-particle transfer operator that is commonly used in the
Interacting Boson-Fermion Model (IBFM), has been derived in the seniority
scheme \cite{olaf}. Although strictly speaking this derivation is only valid
in the vibrational regime, it has been used for deformed nuclei as well.
An alternative method is based on symmetry considerations.
It consists in expressing the single-particle transfer operator
in terms of tensor operators
under the subgroups that appear in the group chain of a dynamical
(super)symmetry \cite{spin6,BI,barea}. The single-particle transfer between
different members of the same supermultiplet provides an important test of
supersymmetries, since it involves the transformation of a boson into a
fermion or vice versa, but it conserves the total number of bosons plus
fermions.

The operators that describe one-proton transfer reactions in the
$U(6/12)_{\nu} \otimes U(6/4)_{\pi}$ supersymmetry are given by \cite{barea}
\begin{eqnarray} 
T_{1,m}^{(\frac{1}{2},\frac{1}{2},-\frac{1}{2}),(\frac{1}{2},\frac{1}{2}),
\frac{3}{2}} &=& -\sqrt{\frac{1}{6}} \left( \tilde{s}_{\pi} \times
a^{\dagger}_{\pi,\frac{3}{2}} \right)^{(\frac{3}{2})}_m
+\sqrt{\frac{5}{6}} \left( \tilde{d}_{\pi} \times
a^{\dagger}_{\pi,\frac{3}{2}} \right)^{(\frac{3}{2})}_m ~,
\nonumber\\
T_{2,m}^{(\frac{3}{2},\frac{1}{2},\frac{1}{2}),(\frac{1}{2},\frac{1}{2}),
\frac{3}{2}} &=& \sqrt{\frac{5}{6}} \left( \tilde{s}_{\pi} \times
a^{\dagger}_{\pi,\frac{3}{2}} \right)^{(\frac{3}{2})}_m
+\sqrt{\frac{1}{6}} \left( \tilde{d}_{\pi} \times
a^{\dagger}_{\pi,\frac{3}{2}} \right)^{(\frac{3}{2})}_m ~.
\label{top1}
\end{eqnarray}
The operators $T_1$ and $T_2$ are, by construction, tensor operators under
$Spin(6)$, $Spin(5)$ and $Spin(3)$ \cite{barea}. The upper indices
$(\sigma_{1},\sigma_{2},\sigma_{3})$, $(\tau_{1},\tau_{2})$, $J$
specify the tensorial properties under $Spin(6)$, $Spin(5)$ and $Spin(3)$.
The use of tensor operators to describe single-particle transfer
reactions in the supersymmetry scheme has the advantage of giving rise to
selection rules and closed expressions for the spectroscopic
factors.

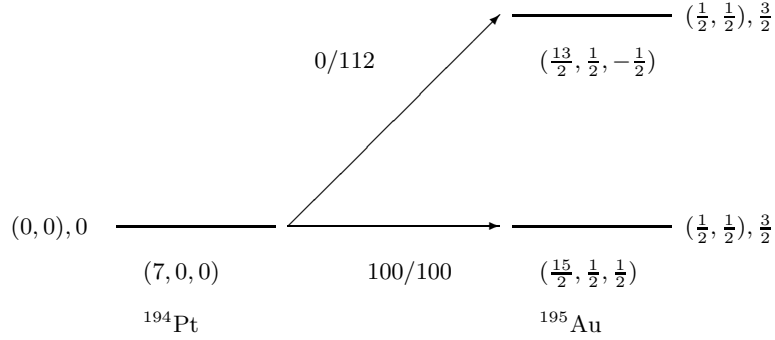
\begin{figure}[t]
\centering
\setlength{\unitlength}{1.0pt}
\begin{picture}(300,160)(0,0)
\thicklines
\put ( 50, 60) {\line(1,0){60}}
\put (200, 60) {\line(1,0){60}}
\put (200,140) {\line(1,0){60}}
\put (125,120) {$0/112$}
\put (145, 40) {$100/100$}
\put ( 60, 40) {$(7,0,0)$}
\put ( 60, 20) {$^{194}$Pt}
\put ( 10, 57) {$(0,0),0$}
\put (210, 40) {$(\frac{15}{2},\frac{1}{2},\frac{1}{2})$}
\put (210, 20) {$^{195}$Au}
\put (265, 57) {$(\frac{1}{2},\frac{1}{2}),\frac{3}{2}$}
\put (210,120) {$(\frac{13}{2},\frac{1}{2},-\frac{1}{2})$}
\put (265,137) {$(\frac{1}{2},\frac{1}{2}),\frac{3}{2}$}
\thinlines
\put (115, 60) {\vector( 1, 0){80}}
\put (115, 60) {\vector( 1, 1){80}}
\end{picture}
\caption[]{\small Allowed one-proton transfer reactions for
$^{194}$Pt $\rightarrow$ $^{195}$Au. The spectroscopic factors
are normalized to 100 for the ground state to ground state
transition for the operators $T_1/T_2$.}
\label{spec1}
\end{figure}

Fig.~\ref{spec1} shows the allowed transitions for the transfer operators
of Eq.~(\ref{top1}) that describe the one-proton transfer from the ground
state $|(N+2,0,0)$, $(0,0)$, $0\rangle$ of the
even-even nucleus $^{194}$Pt to
the even-odd nucleus $^{195}$Au belonging to the supermultiplet
$[N_{\nu}+1\} \otimes [N_{\pi}+1\}_{\pi}$. The number
of bosons $N$ is taken to be the number of bosons in the odd-odd nucleus
$^{196}$Au: $N=N_{\nu}+N_{\pi}$ ($=5$). The operators $T_1$ and
$T_2$ have the same transformation character under $Spin(5)$ and $Spin(3)$,
and therefore can only excite states with
$(\tau_1,\tau_2)=(\frac{1}{2},\frac{1}{2})$ and $J=\frac{3}{2}$. However,
they differ in their $Spin(6)$ selection rules. Whereas $T_1$
can only excite the ground state of the even-odd nucleus with
$(\sigma_1,\sigma_2,\sigma_3)=(N+\frac{3}{2},\frac{1}{2},\frac{1}{2})$,
the operator $T_2$ also allows the transfer to an excited state with
$(N+\frac{1}{2},\frac{1}{2},-\frac{1}{2})$.
The ratio of the intensities is given by \cite{barea}
\begin{eqnarray}
R_1 &=& \frac{I_{\rm gs \rightarrow exc}}{I_{\rm gs \rightarrow gs}}
\;=\; 0 ~,
\nonumber\\
R_2 &=& \frac{I_{\rm gs \rightarrow exc}}{I_{\rm gs \rightarrow gs}}
\;=\; \frac{9(N+1)(N+5)}{4(N+6)^2} ~,
\label{ratios}
\end{eqnarray}
for $T_1$ and $T_2$, respectively. In the case of the one-proton transfer
$^{194}$Pt $\rightarrow$ $^{195}$Au, the second ratio is given by
$R_2=1.12$ ($N=5$).

The available experimental data from the proton stripping reactions
$^{194}$Pt$(\alpha,t)^{195}$Au and $^{194}$Pt$(^{3}$He$,d)^{195}$Au
\cite{munger} shows that the $J=3/2$ ground state of $^{195}$Au is excited
strongly with $C^2S=0.175$, whereas the first excited $J=3/2$ state is
excited weakly with $C^2S=0.019$. In the SUSY scheme, the latter state is
assigned as a member of the ground state band with
$(\tau_1,\tau_2)=(5/2,1/2)$. Therefore the one proton transfer to this
state is forbidden by the $Spin(5)$ selection rule of the tensor operators
of Eq.~(\ref{top1}). The relatively small strength to excited $J=3/2$
states suggests that the operator $T_1$ of Eq.~(\ref{top1}) can be used
to describe the data.

\begin{figure}[t]
\centering
\setlength{\unitlength}{1.0pt}
\begin{picture}(300,160)(0,0)
\thicklines
\put ( 50, 60) {\line(1,0){60}}
\put (200, 60) {\line(1,0){60}}
\put (200,140) {\line(1,0){60}}
\put (125,120) {$0/112$}
\put (145, 40) {$100/100$}
\put ( 60, 40) {$(7,0,0)$}
\put ( 60, 20) {$^{195}$Pt}
\put (  5, 57) {$(0,0),0,\frac{1}{2}$}
\put (210, 40) {$(\frac{15}{2},\frac{1}{2},\frac{1}{2})$}
\put (210, 20) {$^{196}$Au}
\put (265, 57) {$(\frac{1}{2},\frac{1}{2}),\frac{3}{2},L$}
\put (210,120) {$(\frac{13}{2},\frac{1}{2},-\frac{1}{2})$}
\put (265,137) {$(\frac{1}{2},\frac{1}{2}),\frac{3}{2},L$}
\thinlines
\put (115, 60) {\vector( 1, 0){80}}
\put (115, 60) {\vector( 1, 1){80}}
\end{picture}
\caption[]{\small As Fig.~\ref{spec1}, but for
$^{195}$Pt $\rightarrow$ $^{196}$Au.}
\label{spec2}
\end{figure}
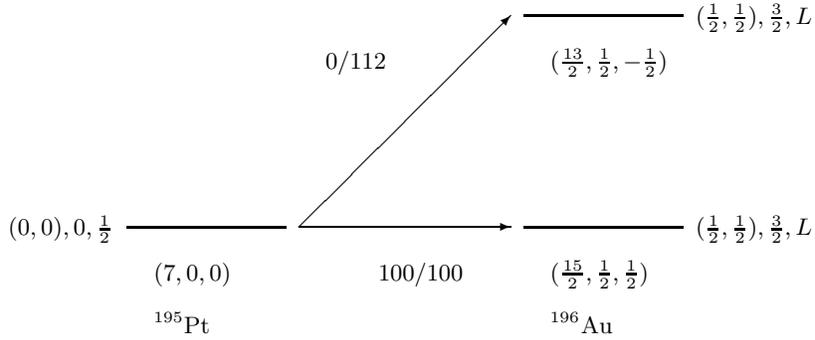

In Fig.~\ref{spec2} we show the allowed transitions for the one-proton
transfer from the ground state $|(N+2,0,0),(0,0),0,\frac{1}{2}\rangle$
of the odd-even nucleus $^{195}$Pt to the odd-odd nucleus $^{196}$Au.
Also in this case, the operator $T_1$ only excites the ground state doublet
of $^{196}$Au with
$(\sigma_1,\sigma_2,\sigma_3)=(N+\frac{3}{2},\frac{1}{2},\frac{1}{2})$,
$(\tau_1,\tau_2)=(\frac{1}{2},\frac{1}{2})$, $J=\frac{3}{2}$ and
$L=J\pm\frac{1}{2}$, whereas $T_2$ also populates the excited state with
$(N+\frac{1}{2},\frac{1}{2},-\frac{1}{2})$. The ratio of the intensities
is the same as for the $^{194}$Pt $\rightarrow$ $^{195}$Au transfer
reaction
\begin{eqnarray}
R_1(^{195}\mbox{Pt} \rightarrow ^{196}\mbox{Au}) &=&
R_1(^{194}\mbox{Pt} \rightarrow ^{195}\mbox{Au}) \;=\; 0 ~,
\nonumber\\
R_2(^{195}\mbox{Pt} \rightarrow ^{196}\mbox{Au}) &=&
R_2(^{194}\mbox{Pt} \rightarrow ^{195}\mbox{Au}) \;=\;
\frac{9(N+1)(N+5)}{4(N+6)^2} ~.
\end{eqnarray}
This is direct consequence of the supersymmetry. Just as the energies
and the electromagnetic transition rates of the supersymmetric quartet of
nuclei were calculated with the same form of the Hamiltonian and the
transition operator, here we have extended this idea to the one-proton
transfer reactions. We find definite predictions for the spectroscopic
factors of the $^{195}$Pt $\rightarrow$ $^{196}$Au transfer reactions,
which can be tested experimentally. To the best of our knowledge, there
are no data available for this reaction.

For the one-neutron transfer reactions there exists a similar situation.
The available experimental data from the neutron stripping reactions
$^{194}$Pt $(d,p)^{195}$Pt \cite{sheline} can be used to determine the
appropriate form of the one-neutron transfer operator \cite{BI}, which
then can be used to predict the spectroscopic factors for the transfer
reaction $^{195}$Au $\rightarrow$ $^{196}$Au. We believe that, as a
consequence of the supersymmetry classification, a number of additional
correlations exist for transfer reactions between different pairs of nuclei.
This would be the first time that such relations are predicted for nuclear
reactions, something which may provide a challenge and motivation for future
experiments.

\subsection{New Experiments}

The great majority of tests carried out for the nuclear supersymmetry
involves one-nucleon transfer experiments such as
$^{197}$Au$(\vec d, t)^{196}$Au and $^{196}$Pt $(\vec d,t)^{195}$Pt
that, in first approximation, are formulated using a transfer
operator of the form $a^\dagger_\nu$. These reactions are very
useful to measure energies, angular momenta and parity of the
residual nucleus. However, they do not test correlations present in the
quartet's wave functions as the case for one-nucleon transfer
reactions inside the supermultiplet (see previous section).
The latter  reactions do provide a direct test of the fermionic
sector (operators $F_{ij}$ and $G_{ij}$ of Eq.~(\ref{fgen}))
of the graded Lie Algebras $U_{\nu}(6/12)$ and $U_{\pi}(6/4)$.

New experimental facilities and detection 
techniques \cite{tres,pt195,au196,wirth} offer a unique opportunity for
analyzing the supersymmetry classification in greater
detail \cite{graw}. In reference \cite{barea} we pointed out a
symmetry route for the theoretical analysis of such reactions, via
the use of tensor operators of the algebras and superalgebras.  An
alternative route is the use of a semi-microscopic approach where
projection techniques starting from the original  nucleon pairs
lead to specific forms for  the operators \cite{olaf,quince}
which, however, are only strictly  valid in the generalized
seniority regime \cite{dieciseis}. The former and latter
routes may be related by a consistent-operator approach, where the
Hamiltonian exchange operators are made to be consistent with
the one-nucleon transfer operator implying that the exchange term
in the boson-fermion Hamiltonian can be viewed as an internal
exchange reaction  among the nucleon and the nucleon pairs. In
addition to these experiments, ongoing research explores the
possibility of testing SUSY through new transfer reactions. The
two-nucleon transfer $(\alpha,\vec d)$ and $(\vec d,\alpha)$
reactions probe neutron-proton correlations in the nuclear wave
function and constitutes a very stringent test of the supersymmetry
classification.

In particular, the $^{194}$Pt$(\alpha,\vec d)^{196}$Au reaction
involves nuclei belonging to the same supermultiplet. Therefore
this process can be described by a combination of the fermionic
generators the superalgebra (see Eq.~(\ref{fgen})).
Likewise, the reaction $^{195}$Pt$(^3$He$,t)^{195}$Au is expressible in
terms of the fermionic operators which, in this case, is
associated to the beta-decay operator \cite{dieciocho}. These
reactions and their relation to single-nucleon transfer
experiments raise the exciting possibility of testing direct
correlations among transfer reaction spectroscopic factors in
different nuclei, predicted by the supersymmetric classification
of the magic quartet. A preliminary report on these analyses was 
presented in Ref.~\cite{diez}.

\subsection{SUSY without Dynamical Symmetry}

\begin{figure}[t]
\centering
$\begin{array}{lr}
\includegraphics[height=10cm,width=5.5cm]{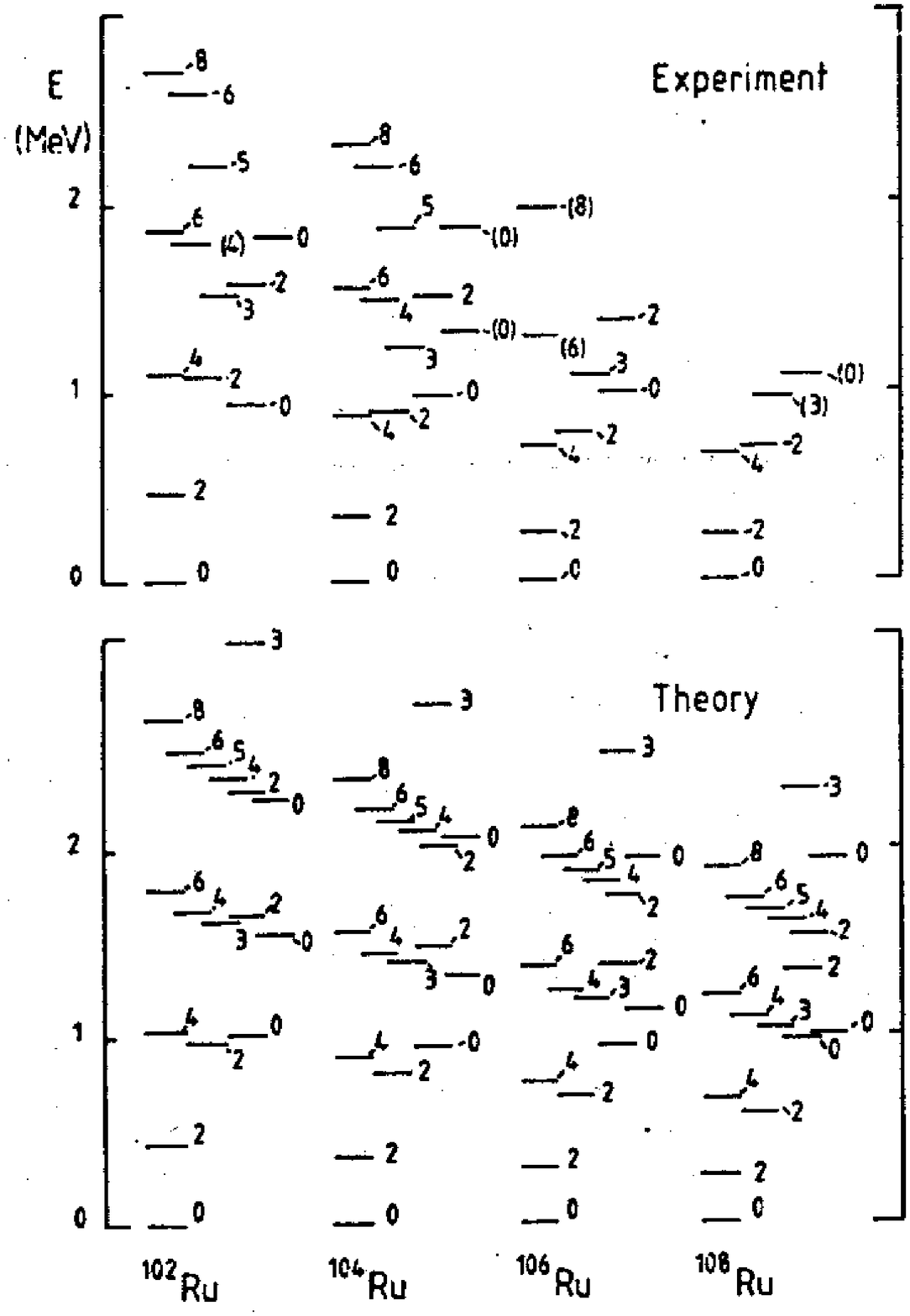} &
\includegraphics[height=10cm,width=5.5cm]{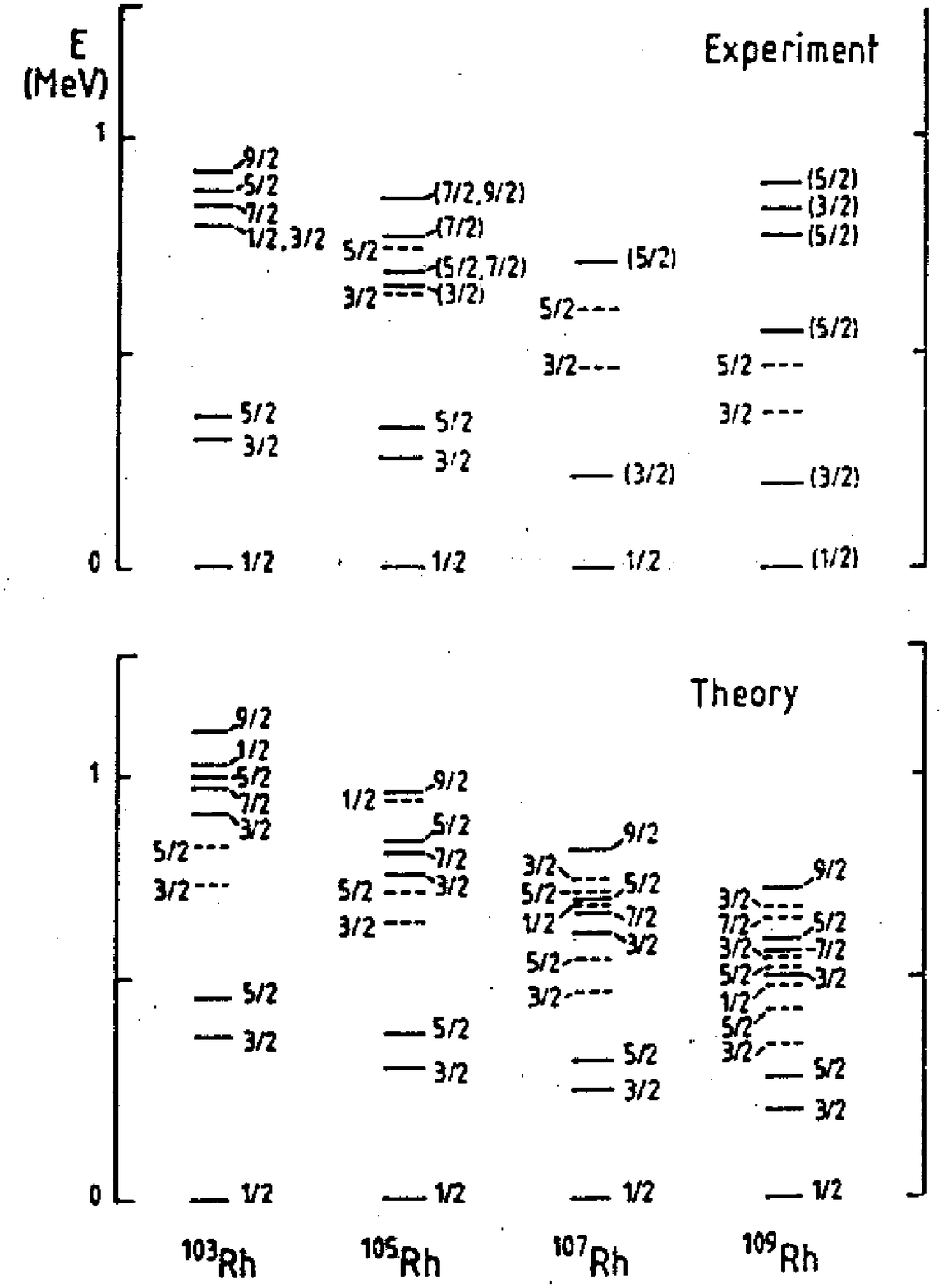}
\end{array}$
\caption{\small Experimental and calculated positive-parity states in
$^{102-108}$Ru and negative-parity states in $^{103-109}$Rh
\protect\cite{once}.}\label{ru-rh}
\end{figure}

The concept of dynamical algebra (not to be confused with that of 
dynamical symmetry) implies a generalization of the concept of 
symmetry algebra, as explained in Section 2.2. If $G$ is the 
dynamical algebra of a system, \underbar{all} physical states 
considered belong to a single irreducible representation (IR) of 
$G$. (In a symmetry algebra, in contrast, each set of degenerate 
states of the system is associated to an IR).  The best known 
examples of a dynamical algebra are perhaps  $SO(4,2)$ for the 
hydrogen atom and the $U(6)$ IBM algebra for even-even nuclei. A 
consequence of having a dynamical algebra associated to a system 
is that all sates can be reached using the algebra's generators 
or, equivalently, all physical operators can be expressed in terms 
of these operators \cite{nueve}.  Naturally, the same Hamiltonian 
and the same transition operators are employed for all states in 
the system.  To further clarify this point, it is certainly true 
that a single $H$ and a single set of  operators are associated to 
a given even-even nucleus in the IBM framework, expressed in terms 
of the $U(6)$ (dynamical algebra) generators.  It doesn't matter 
whether this Hamiltonian can be expressed or not in terms of the 
generators of a single chain of groups (a dynamical symmetry). 

\begin{figure}[t]
\centering
\includegraphics[height=12cm,width=6cm]{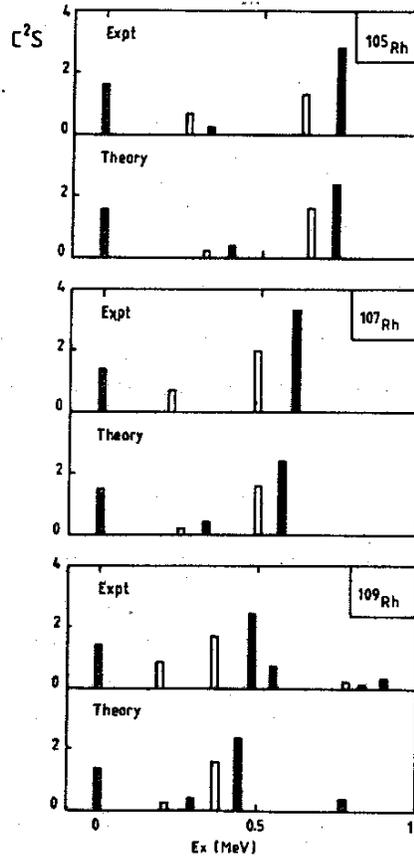} 
\caption{\small Experimental and calculated spectroscopic factors 
in Rh isotopes \protect\cite{once}.} 
\label{rhspec}
\end{figure}

In the same fashion, if we now consider $U(6/12)$ to be the 
dynamical algebra for the pair of nuclei $^{194}$Pt-$^{195}$Pt, it 
follows that the same $H$ and operators (including in this case 
the transfer operators that connect states in the different 
nuclei) should apply to all states.  It also follows that no 
restriction should be imposed on the form of $H$, except that it 
must be a function of the generators of $U(6/12)$ (the enveloping 
space associated to it).  It should be clear that the concept of 
supersymmetry does not require the existence of a particular 
dynamical symmetry.  Extending these ideas to the neutron-proton space 
of IBM-2 we can say that SUSY is equivalent to requiring that a 
product of the form
\begin{equation}
U_{\nu}(6/\Omega_{\nu}) \otimes U_{\pi}(6/\Omega_{\pi}) ~,
\end{equation}
plays the role of dynamical (super)algebra for a quartet of even-even,
even-odd, odd-even and odd-odd nuclei. Having said that, it should be
stated that the dynamical supersymmetry has the distinct advantage of
immediately suggesting the form of the quartet's Hamiltonian
and operators, while the general statement made above does not
provide a general recipe. For some particular cases, however,
this can be done in a straightforward way. In
reference~\cite{once}, for example, the $U(6/12)$ supersymmetry
(without imposing any of the three dynamical IBM symmetries) was
successfully tested for the Ru and Rh isotopes. In that case a
combination of $U^{BF}(5)$ and $SO^{BF}(6)$ symmetries was shown
to give an excellent description of the data, as shown in 
Figs.~\ref{ru-rh} and~\ref{rhspec}.

An immediate consequence of this proposal is that it opens up the
possibility of testing SUSY in other nuclear regions, since
dynamical symmetries are very scarce and have severely limited the
study of nuclear supersymmetry.

\section{Summary and Conclusions}

In these lecture notes we have discussed different aspects
of supersymmetry in nuclear physics. We have attempted to give a
general overview of the subject, starting from the fundamental
concepts in group theory and Lie algebras, which are the required 
mathematical framework for this phenomenon.

The recent measurements of the spectroscopic properties of the
odd-odd nucleus $^{196}$Au have rekindled the interest in nuclear
supersymmetry, as has been discussed in some detail. The available
data on the spectroscopy of the quartet of nuclei $^{194}$Pt,
$^{195}$Au, $^{195}$Pt and $^{196}$Au  can, to a good
approximation, be described in terms of the $U(6/4)_{\pi}\otimes
U(6/12)_{\nu}$ supersymmetry. However, there is a still another
important set of experiments which can further test the
predictions of the supersymmetry scheme. These involve transfer
reactions between nuclei belonging to the same supermultiplet, in
particular between the even-odd (odd-even) and odd-odd members of
the supersymmetric quartet. Theoretically, these transfers are
described by the supersymmetric generators which change a boson
into a fermion, or vice versa.

We have discussed the example of proton transfer between the SUSY
partners: $^{194}$Pt $\rightarrow$ $^{195}$Au and $^{195}$Pt
$\rightarrow$ $^{196}$Au. The supersymmetry implies strong
correlations for the spectroscopic factors of these two reactions
which can be tested experimentally. A similar set of relations
can be derived for the one-neutron transfer reactions $^{194}$Pt
$\leftrightarrow$ $^{195}$Pt and $^{195}$Au $\leftrightarrow$
$^{196}$Au. Another interesting extension of supersymmetry
concerns the recently measured two-nucleon transfer reaction
$^{194}$Pt$(\alpha,d)^{196}$Au \cite{graw}, in which a
neutron-proton pair is transferred to the target nucleus. This
reaction presents a very sensitive test of the wave functions,
since it provides a measure of the correlation within the
transferred neutron-proton pair. Whether it is possible to
describe this process by a transfer operator that is correlated by
SUSY to that of the one-proton and one-neutron transfer reactions
is an open question.In these lecture notes we have also argued that
n-Susy can in principle be generalized to encompass transitional
nuclei, that is, that do not correspond to dynamical symmetries.

\begin{figure}[t]
\centering
\includegraphics[height=11.5cm,width=8.5cm]{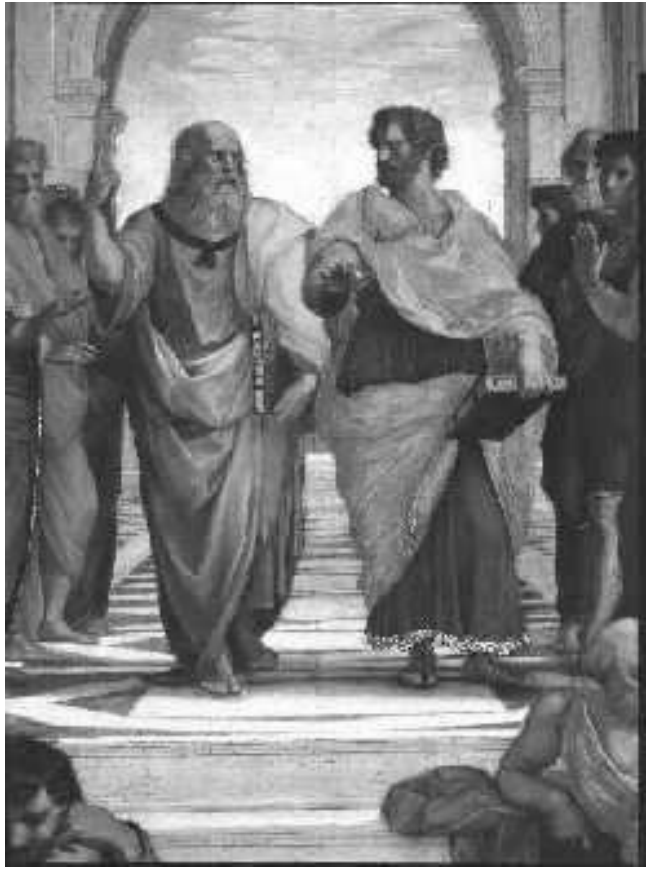}
\caption{\small Detail of ``The School of Athens'' (Plato on the left and
Aristoteles on the right), by Rafael.}
\label{athens}
\end{figure}

In conclusion, we have reviewed the current status of nuclear
supersymmetry and considered diverse extensions that are
currently being investigated.  We have emphasized the need for
further experiments taking advantage of new experimental
capabilities \cite{tres,pt195,au196}, suggesting that particular
attention be paid to one- and two-nucleon transfer reactions
between the SUSY partners $^{194}$Pt, $^{195}$Au, $^{195}$Pt and
$^{196}$Au, since such experiments provide the most stringent
tests of nuclear supersymmetry. It remains to be seen whether the
correlations predicted by n-SUSY are indeed verified by new
experiments and whether these correlations can be truly extended
to mixed-symmetry regions of the nuclear table. If this is the
case, nuclear supersymmetry may yet provide a powerful unifying
scheme for atomic nuclei, thus becoming a particularly striking
example of the combination of the Platonic ideal of symmetry with
the down-to-earth Aristotelic ability to recognize complex
patterns in Nature.

\section*{Acknowledgments}

We are grateful to C. Alonso, J. Arias, G. Graw, J. Jolie, P. Van Isacker 
and H.-F. Wirth for interesting discussions. We are particularly indebted to 
R. Lemus for his artistry and generosity in designing 
Fig.~\ref{aztec}. This paper was supported in part by Conacyt, Mexico.

\printindex
\end{document}

%% file: mac.tex
\def\bx{{\bar x}}

\def\bfD{{\bf D}}
\def\bfG{{\bf G}}
\def\bfM{{\bf M}}
\def\bfR{{\bf R}}
\def\bfS{{\bf S}}
\def\bfT{{\bf T}}
\def\bfU{{\bf U}}


\def\bfmh{{\vec{h}}}
\def\bfml{{\bf l}}
\def\bfmL{{\bf L}}
\def\bfmp{{\bf p}}
\def\bfmr{{\bf r}}
\def\bfmR{{\bf R}}
\def\bfmx{{\bf x}}
\def\bfom{{\vec{\omega}}}
\def\bfps{{\vec{\psi}}}

\def\bfoms{{\vec{\omega}}}

\def\cA{{\cal A}}
\def\cB{{\cal B}}
\def\cC{{\cal C}}
\def\cD{{\cal D}}
\def\cG{{\cal G}}
\def\cN{{\cal N}}
\def\cO{{\cal O}}
\def\cP{{\cal P}}
\def\cdP{{\cal P}^\dagger}
\def\ctG{\tilde{\cal G}}
\def\ctP{\tilde{\cal P}}
\def\cQ{{\cal Q}}
\def\cY{{\cal Y}}

\def\da{a^\dagger}
\def\db{b^\dagger}
\def\dd{d^\dagger}
\def\dh{h^\dagger}
\def\dpz{p^\dagger}
\def\ds{s^\dagger}
\def\dt{t^\dagger}

\def\hA{{\hat A}}
\def\hB{{\hat B}}
\def\hD{{\hat D}}
\def\he{{\hat e}}
\def\hE{{\hat E}}
\def\hF{{\hat F}}
\def\hg{{\hat g}}
\def\hG{{\hat G}}
\def\hH{{\hat H}}
\def\hj{{\hat\jmath}}
\def\hJ{{\hat J}}
\def\hk{{\hat k}}
\def\hK{{\hat K}}
\def\hl{{\hat l}}
\def\hL{{\hat L}}
\def\hm{{\hat m}}
\def\hM{{\hat M}}
\def\hn{{\hat n}}
\def\hN{{\hat N}}
\def\hO{{\hat O}}
\def\hP{{\hat P}}
\def\hQ{{\hat Q}}
\def\hR{{\hat R}}
\def\hS{{\hat S}}
\def\hT{{\hat T}}
\def\hU{{\hat U}}
\def\hV{{\hat V}}
\def\hX{{\hat X}}

\def\hcN{{\hat{\cal N}}}

\def\oh{\bar{h}}
\def\oO{\overline{\rm O}}
\def\osigma{\bar{\sigma}}
\def\oSO{\overline{\rm SO}}
\def\oSU{\overline{\rm SU}}
\def\oU{\overline{\rm U}}

\def\rb{{\rm b}}
\def\rbent{{\rm bent}}
\def\rB{{\rm B}}
\def\rBF{{\rm BF}}
\def\rd{{\rm d}}
\def\re{{\rm e}}
\def\rE{{\rm E}}
\def\reven{{\rm even}}
\def\rf{{\rm f}}
\def\rF{{\rm F}}
\def\rg{{\rm g}}
\def\rhw{{\rm hw}}
\def\ri{{\rm i}}
\def\rI{{\rm I}}
\def\rII{{\rm II}}
\def\rIII{{\rm III}}
\def\rJ{{\rm J}}
\def\rlin{{\rm lin}}
\def\rM{{\rm M}}
\def\rN{{\rm N}}
\def\rns{{\rm ns}}
\def\ro{{\rm o}}
\def\rO{{\rm O}}
\def\rodd{{\rm odd}}
\def\rp{{\rm p}}
\def\rr{{\rm r}}
\def\rR{{\rm R}}
\def\rrv{{\rm rv}}
\def\rrig{{\rm rig}}
\def\rrve{{\rm rv-e}}
\def\rs{{\rm s}}
\def\rsd{{\rm sd}}
\def\rSO{{\rm SO}}
\def\rSp{{\rm Sp}}
\def\rst{{\rm st}}
\def\rSU{{\rm SU}}
\def\rt{{\rm t}}
\def\ru{{\rm u}}
\def\rU{{\rm U}}
\def\rv{{\rm v}}
\def\rx{{\rm x}}

\def\ta{{\tilde a}}
\def\tb{{\tilde b}}
\def\td{{\tilde d}}
\def\tep{{\tilde\epsilon}}
\def\tK{{\tilde K}}
\def\tl{{\tilde l}}
\def\tn{{\tilde n}}
\def\tp{{\tilde p}}
\def\ts{{\tilde s}}
\def\tSO{{\widetilde\rSO}}
\def\tt{{\tilde t}}

\def\seq{\!=\!}
\def\sitem{\item\vspace{-3mm}}
\def\smin{\!-\!}
\def\smp{\!\mp\!}
\def\sparallel{\!\parallel\!}
\def\splus{\!+\!}
\def\spm{\!\pm\!}
\def\ssmin{\!\!-\!\!}
\def\ssplus{\!\!+\!\!}
\def\stimes{\!\times\!}

\def\mink{\!-\!}
\def\mpk{\!\mp\!}
\def\plusk{\!+\!}
\def\pmk{\!\pm\!}

\def\dscal{\!:\!}
\def\scal{\cdot}

\def\tc{\!\times\!}

\def\tf#1#2{
\raise0.40ex\hbox{$\scriptstyle#1$}
\raise0.20ex\hbox{$\scriptstyle/$}
\raise0.00ex\hbox{$\scriptstyle#2$}}
\def\to#1#2{{\textstyle{#1\over#2}}}
\def\spin#1{
\raise0.40ex\hbox{$\scriptstyle#1$}
\raise0.20ex\hbox{$\scriptstyle/$}
\raise0.00ex\hbox{$\scriptstyle2$}}

\def\pder#1{{{\partial}\over{\partial#1}}}
\def\pderm#1#2{{{\partial^{#1}}\over{\partial{#2}^{#1}}}}
\def\der#1{{{d}\over{d#1}}}
\def\derm#1#2{{{d^{#1}}\over{d{#2}^{#1}}}}

\def\bin#1#2{\biggl(\!\begin{array}{c}#1\\#2\end{array}\!\biggr)}
\def\pthree#1#2#3#4#5#6{\biggl(\!\begin{array}{ccc}
#1&#2&#3\\#4&#5&#6
\end{array}\!\biggr)}
\def\racah#1#2#3#4#5#6{\biggl\{\!\begin{array}{ccc}
#1&#2&#3\\#4&#5&#6
\end{array}\!\biggr\}}
\def\ninej#1#2#3#4#5#6#7#8#9{\left\{\!\begin{array}{ccc}
#1&#2&#3\\#4&#5&#6\\#7&#8&#9
\end{array}\!\right\}}
\def\xninej#1#2#3#4#5#6#7#8#9{\left[\!\begin{array}{ccc}
#1&#2&#3\\#4&#5&#6\\#7&#8&#9
\end{array}\!\right]}
\def\coupt#1#2#3#4#5#6{\langle#1#2\otimes#3#4\vert#5#6\rangle}
\def\coup#1#2#3#4#5#6{\left\langle\!\begin{array}{cc|c}
#1&#3&#5\\#2&#4&#6
\end{array}\!\right\rangle}

\def\tauv{{v}}

\def\dsum#1#2{{\displaystyle \sum_{{\scriptstyle #1}\atop{\scriptstyle #2}}}}
\def\sumd#1{{\displaystyle \sum_{#1}}}

\def\youngonerow#1{\begin{array}{l}
\overbrace{\fox\fox\cdots\fox}^{#1}
\end{array}}
\def\youngonerowone#1{\begin{array}{l}
\overbrace{\fox\fox\fox\cdots\fox}^{#1}\\
\fox
\end{array}}
\def\youngtworow#1#2{\begin{array}{l}
\overbrace{\fox\fox\fox\cdots\fox}^{#1}\\
\overbrace{\fox\fox\cdots\fox}^{#2}
\end{array}}
\def\youngnrow#1#2#3{\begin{array}{l}
\overbrace{\fox\fox\fox\cdots\fox}^{#1}\\
\overbrace{\fox\fox\cdots\fox}^{#2}\\
\vdots\\
\overbrace{\fox\cdots\fox}^{#3}
\end{array}}
\def\youngonecolumn#1{\left.\begin{array}{c}
\fox\\\fox\\\vdots\\\fox
\end{array}
\right\}#1}
\def\youngtwocolumn#1#2{\left.\begin{array}{c}
\fox\\\fox\\\fox\\\vdots\\\fox
\end{array}
\right\}#1
\left.\begin{array}{c}
\fox\\\fox\\\vdots\\\fox
\end{array}
\right\}#2}

\def\youngone{\begin{array}{c}
\fox
\end{array}}
\def\youngoneone{\begin{array}{c}
\fox\\\fox
\end{array}}
\def\youngoneoneone{\begin{array}{l}
\fox\\\fox\\\fox
\end{array}}
\def\youngoneoneoneone{\begin{array}{l}
\fox\\\fox\\\fox\\\fox
\end{array}}
\def\youngtwo{\fox\fox}
\def\youngtwooneone{\begin{array}{l}
\fox\fox\\\fox\\\fox
\end{array}}
\def\youngtwotwo{\begin{array}{l}
\fox\fox\\\fox\fox
\end{array}}
\def\youngtwotwooneone{\begin{array}{l}
\fox\fox\\\fox\fox\\\fox\\\fox
\end{array}}
\def\youngtwotwotwo{\begin{array}{l}
\fox\fox\\\fox\fox\\\fox\fox
\end{array}}

\def\fox{\mbox{\large$\Box$}}
\def\foxa{\fox\kern-0.65em\raise0.55ex\hbox{$\scriptstyle{a}$}\kern0.27em}
\def\foxb{\fox\kern-0.63em\raise0.40ex\hbox{$\scriptstyle{b}$}\kern0.25em}

\def\chainspace{\arraycolsep=0.15em}

\def\References{
\section*{References}
\markright{\bf References}}
\def\Summary{
\section*{Summary}
\markright{\bf Summary}}

\def\refboo#1#2#3#4#5
{#1, {\it#2}, #3, #4, #5}
\def\refart#1#2#3#4#5#6
{#1, ``#2,'' #3 {\bf#4} (#5) #6}
\def\refarts#1#2#3#4#5
{#2 {\bf#3} (#4) #5}
\def\refcon#1#2#3#4#5#6#7#8
{#1, ``#2,'' in {\it#3}, edited by #4, #5, #6, #7, p.~#8}
\def\refthe#1#2#3#4
{#1, {\it#2}, doctoral dissertation, #3, #4}

\def\beq{\begin{equation}}
\def\eeq{\end{equation}}
\def\beqn{\begin{equation}} 
\def\eeqn{\end{equation}} 
\def\beqa{\begin{eqnarray}}
\def\eeqa{\end{eqnarray}}
\def\beqan{\begin{eqnarray}} 
\def\eeqan{\end{eqnarray}} 
\def\non{\nonumber\\}

\def\dim{{\rm dim}}

\hyphenation{
ana-logy
di-men-sion di-men-sions di-men-sion-al
ei-gen-func-tion ei-gen-func-tions
ei-gen-state ei-gen-states
ei-gen-val-ue ei-gen-val-ues
in-fini-tesi-mal
mol-ecule mol-ecules mol-ecu-lar
quad-rat-ic
ro-ta-tion ro-ta-tions ro-ta-tion-al
single
sym-me-try sym-me-tries sym-met-ric anti-sym-met-ric
su-per-sym-me-try su-per-sym-me-tries su-per-sym-met-ric
vi-bra-tion vi-bra-tions vi-bra-tion-al
wheth-er}